\documentclass[aps,pre,twocolumn,showpacs,showkeys,preprintnumbers,amsmath,amssymb]{revtex4-1}
\usepackage{amscd,graphicx,color,colordvi,bm,amsfonts,stmaryrd}
\usepackage{dsfont}
\usepackage{hyperref}

\renewcommand{\exp}[1]{{\text e}^{#1}}
\newcommand{\avg}[1]{\left\langle #1 \right\rangle}

\newcommand{\pdif}[2]{\frac{\partial #1}{\partial #2}}
\newcommand{\dif}[2]{\frac{\D #1}{\D #2}}
\renewcommand{\vec}[1]{\mathbf{#1}}

\newcommand{\mat}[1]{\mathbf{#1}}
\renewcommand{\avg}[1]{\langle #1 \rangle}

\newcommand{\D}{\text{d}}

\newcommand{\kB}{k_\text{B}}
\newcommand{\tf}{t_\text{f}}

\newcommand{\sm}{\sig_\text{inf}}
\newcommand{\sM}{\sig_\text{sup}}
\newcommand{\smss}{\sm^\text{s.s.}}

\newcommand{\ds}{s_\text{tot}}

\newcommand{\sig}{s}
\newcommand{\dS}{\avg{\ds}}

\newcommand{\dSM}{S_\text{sup}}
\newcommand{\dSMss}{\dSM^\text{\,s.s.}}

\newcommand{\dsa}{s^\text{a}}
\newcommand{\dsna}{s^\text{na}}
\newcommand{\dsam}{\sigma^\text{a}_\text{m}}
\newcommand{\dsaM}{\sigma^\text{a}_\text{M}}
\newcommand{\dsnam}{\sigma^\text{na}_\text{m}}
\newcommand{\dsnaM}{\sigma^\text{na}_\text{M}}
\newcommand{\Smax}{S_\text{sup}}

\newcommand{\rvec}{\vec{r}}
\newcommand{\xvec}{\vec{x}}
\newcommand{\uvec}{\vec{u}}
\newcommand{\vvec}{\vec{v}}
\newcommand{\vcup}{\tilde{\vec{v}}}
\newcommand{\g}{g}
\newcommand{\gdual}{g^*}

\newcommand{\Sdual}{S^{*}}

\newcommand{\Atraj}{\vec{x}_\vec{i}}
\newcommand{\Afam}{\vec{x}_\vec{i}(\delta t)}
\newcommand{\PF}{\mathcal{P}}

\newcommand{\PB}{\mathcal{P}^\text{B}}
\newcommand{\Pp}{\mathcal{P}'}

\newcommand{\Pforward}[2]{\PF_{#1}(#2)}

\newcommand{\Pbackward}[2]{\PB_{#1}(#2)}

\newcommand{\WB}[2]{W^\text{B}_{ #1 \shortleftarrow #2 }}

\newcommand{\rB}{r^\text{B}}

\newcommand{\K}[2]{K_{ #1 \shortleftarrow #2 }}
\newcommand{\rK}{r^K}

\newcommand{\Kmat}{\mat{K}}
\newcommand{\PK}{\mathcal{P}^\dagger}

\newcommand{\PKforward}[2]{\PK_{#1}(#2)}

\newcommand{\Ktraj}{\vec{x}_{\vec{i}^\dagger}}
\newcommand{\Btraj}{\vec{x}_{\vec{i}^\text{B}}}

\newcommand{\Kfam}{\vec{x}_{\vec{i}^\dagger}(\delta t)}

\begin{document}


\title{%
  Upper bound for the average entropy production based on stochastic entropy extrema
}

\author{Surachate Limkumnerd}
\email{surachate.l@chula.ac.th}
\affiliation{%
Department of Physics, Faculty of Science, Chulalongkorn University, Phayathai Rd., Patumwan, Bangkok 10330, Thailand \\
Research Center in Thin Film Physics, Thailand Center of Excellence in Physics, CHE, 328 Si Ayutthaya Rd., Bangkok 10400, Thailand
}

\date{\today}

\begin{abstract}
The second law of thermodynamics, which asserts the non-negativity of the average total entropy production of a combined system and its environment, is a direct consequence of applying Jensen's inequality to a fluctuation relation. It is also possible, through this inequality, to determine an upper bound of the average total entropy production based on the entropies along the most extreme stochastic trajectories. In this work, we construct an upper bound inequality of the average of a convex function over a domain whose average is known. When applied to the various fluctuation relations, the upper bounds of the average total entropy production are established. Finally, by employing the result of Neri, Rold{\'a}n, and J{\"ulicher} [Phys. Rev. X 7, 011019 (2017)], we are able to show that the average total entropy production is bounded only by the total entropy production supremum, and vice versa, for a general non-equilibrium stationary system.
\end{abstract}

\keywords{second law of thermodynamics, fluctuation relations, convex optimization}

\maketitle

\section{Introduction}\label{s:intro}

Statistical mechanics underwent a major development in the early 1990s, and this gave rise to what were to be collectively known as fluctuation relations. It had long been observed that systems with few degrees of freedom are especially susceptible to fluctuations and seemingly behave as if they violate the second law of thermodynamics. These violations are however quantifiable. 
Evans et al.~\cite{EvanCoheMorr93,*EvanSear94,*GallCohe95a} were able to calculate the relative likeliness between a regular trajectory of events and the corresponding reversed trajectory that violates the second law in a strongly sheared fluid system. Independently, Jarzynski~\cite{Jarz97PRL,*Jarz97PRE} and Crooks~\cite{Crooks98,*Crooks99,*Crooks00} found a relationship that ties together the amount of time-dependent work during a non-equilibrium process and the change in the system's free energy. All these relations are valid for an arbitrarily long time, and arbitrarily far from equilibrium. Over the following decades, fluctuation relations were generalized and put onto a more solid mathematical foundation~\cite{ChetFalkGawe08,*GarcLecoKoltDomi12}, and they were experimentally verified in several systems, such as the motion of colloidal particles in a harmonic trap~\cite{WangSeviMittSear02,*CarbWillWangSevi04}, an electrical dipole driven out of equilibrium~\cite{GarnCili05}, and the unfolding and refolding of a small RNA hairpin~\cite{CollRitoJarzSmit05}.

In essence, fluctuation relations can be understood simply as a change of variables in probability space. Let probability measure $\PF$ on a measurable space ($\chi,\Sigma$) be absolutely continuous with respect to another measure $\Pp$ on the same space. One can then define random variable $\ds$ such that
\begin{equation}\label{E:TFT} 
	\exp{-\ds} = \D \Pp/\D \PF.
\end{equation}
Here $\D \Pp/\D \PF$ is formally known as the Radon-Nikod\'ym derivative which describes the rate of change of probability in one measure with respect to the other~\cite{Klenke13}. It follows immediately that
\begin{equation}\label{E:IFT}
	1 = \int_\chi \,\D\Pp = \int_\chi \mathrm{exp}(-\ds)\,\D\PF = \avg{\mathrm{exp}(-\ds)},
\end{equation}
where $\avg{\cdot}$ is used to denote the average over the forward measure.
Eq.~(\ref{E:TFT}) and (\ref{E:IFT}) are known respectively as detailed and integral fluctuation theorems. 
The interested reader is referred to Ref.~\onlinecite{Maes99} and \onlinecite{ChetGupt11} for excellent overviews of the mathematical concepts of  fluctuation theorems. 
At this stage these relations are devoid of any physical meanings. It was shown~\cite{Seif05} that if $\PF$ represents the probability (density) that a forward trajectory $x(t)$ is observed during $0\le t\le \tf$ according to some protocol $\lambda_t$, and $\Pp$ is the probability of observing the corresponding backward trajectory $x^\dagger(t) = x(\tf-t)$ with time-reversed protocol $\lambda^\dagger_t = \lambda_{\tf-t}$, then $\ds$ denotes the entropy production of the combined system and environment along the trajectory during that duration. (See Appendix~\ref{sec:entropyIsBounded} for a mathematical construction and an example of $\PF$, $\Pp$ and $\ds$ in the case of an inhomogeneous Markov jump process.) It is important to note that the initial state of the backward trajectory is taken to be the final state of the forward trajectory, and vice versa. Other choices of protocols and boundary values yield other types of entropies. 
For example, Esposito and Van den Broeck~\cite{EspoBroe10} recently showed that the trajectory-wise total entropy production can be split into two pieces, the adiabatic ($\dsa$) and the nonadiabatic ($\dsna$) pieces, $\ds = \dsa+\dsna$, which separately satisfy fluctuation relations (\ref{E:TFT}) and (\ref{E:IFT}). 
For a comprehensive review of various boundary conditions and the associated entropies, see Ref.~\onlinecite{Seif12}. 

It should be emphasized that $\ds$ is trajectory dependent, and bounded.  (See proof in Appendix~\ref{sec:entropyIsBounded} after Eq.~(\ref{E:entropydef}).) 
We shall denote the supremum (infimum) of the total entropy production at time $\tf$ over all sampling trajectories by $\sM$ ($\sm$). 
Since the exponential function is convex, an application of Jensen's inequality to (\ref{E:IFT}) implies the second law inequality. In other words, $1= \avg{\rm{exp}(-\ds)} \ge \rm{exp}(-\avg{\ds})$ implies that $\avg{\ds} \ge 0$. The equality is reached when the system is 
in equilibrium,
otherwise the system and the environment must produce net positive total entropy on average. Surely there must exist an upper limit to this number that is below $\sM$.
The question is, thus, if nothing else but the two limits $\sM$ and $\sm$ are known, can we place an upper bound of $\dS$ based on them?

The paper is organized as follows. The mathematical structures will be presented in Sec.~\ref{sec:math} where we construct the upper-bound inequality for the single- and multiple-parameter cases. These bounds are applicable to any convex function. 
We shall apply it to obtain the upper bound of the total entropy as a function of adiabatic and nonadiabatic entropy productions along extremal trajectories (Sec.~\ref{sec:FlucRel}), and to show the relationships between the bounds of the average entropy production and the entropy production supremum in the non-equilibrium stationary state case (Sec.~\ref{s:RelnBounds}). Concluding remarks are given in Sec.~\ref{sec:conclusion}. Interpretations of entropy production for jump processes and detailed calculations are provided in Appendix~\ref{sec:entropyIsBounded} to \ref{sec:AppB}.

\section{Mathematical construction of the upper bound}\label{sec:math}

\begin{figure}[htb]
	\includegraphics[width=.47\textwidth]{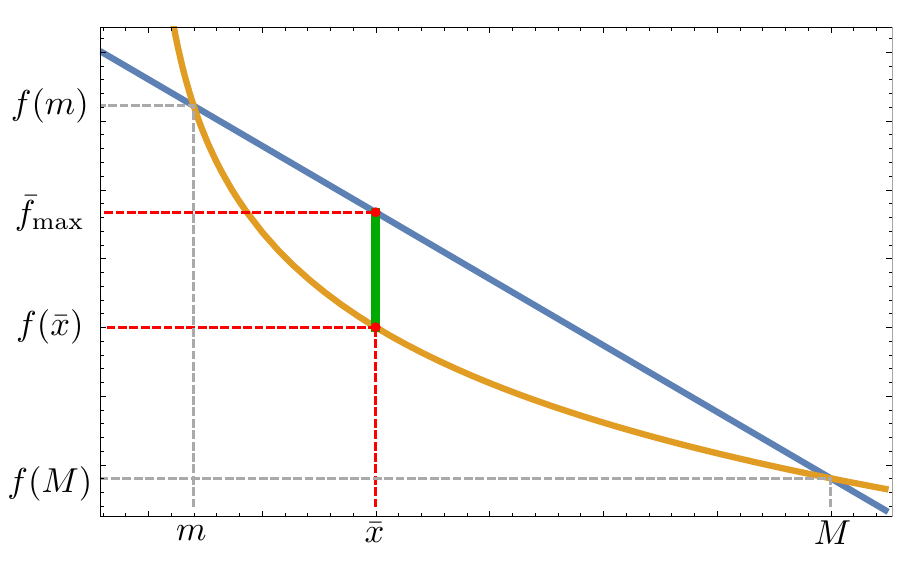}
	\caption{(Color online) By definition a convex function $f(x)$ (yellow curve) must lie below a secant line (blue line) connecting any two points on the curve. Denote these points by $(m,f(m))$ and $(M,f(M))$. Let $\bar{x}$ be the average of random variable $x$. Inequality (\ref{E:IneqProof}) states that the average $\overline{f(x)}$ must be bounded from above by a point on the secant line at $\bar{x}$. This average is also bounded from below according to $f(\bar{x}) \le \overline{f(x)}$ by the direct application of Jensen's inequality. The green line indicates the region of possible values of $\overline{f(x)}$.}
	\label{fig:Kanto}
\end{figure}

Before venturing into the general derivation of computing the upper bound of a convex function over a domain with a known average, we present here the simplest working case. Suppose $x$ is a positive random variable such that $0<m\le x \le M <\infty$ almost surely, and $f$ is a convex function on the domain of $x$. We can parametrize $x$ by its bounds according to
$ x = \lambda_x M + (1-\lambda_x)m$,
where $\lambda_x = (x-m)/(M-m) \in [0,1]$. Due to the convexity of $f(x)$,
\begin{equation}\label{E:beforeProof}
	f(x) \le \lambda_x f(M) + (1-\lambda_x)f(m).
\end{equation}
Since $\lambda_x$ is linear in $x$, its average over all $x$ is simply $\overline{\lambda_x} = \lambda_{\bar{x}}$. Here $\bar{x}$ is the average value of $x$, and is presumed to be known. Note also that this average could be over all possible outcomes (ensemble average) or over time. Taking the average of (\ref{E:beforeProof}) gives
\begin{equation}\label{E:IneqProof}
\begin{split}
	\overline{f(x)} &\le \overline{\lambda_x f(M) + (1-\lambda_x)f(m) } \\
	&= \lambda_{\bar{x}} f(M) + (1-\lambda_{\bar{x}})f(m) \equiv \bar{f}_\text{sup},
\end{split}
\end{equation}
The equality is reached if and only if $m=M$. Fig.~\ref{fig:Kanto} gives a graphical interpretation of this inequality.

In our case, $x = \exp{-\ds}$, $m = \exp{-\sM}$, $M = \exp{-\sm}$, $f(x) = -\ln x$, and according to (\ref{E:IFT}), $\bar{x}=1$. By taking the average to be over the ensemble of trajectories, 
the upper bound for the entropy production is obtained:
\begin{equation}\label{E:upperbounds}
	0 \le \dS \le \Smax \equiv \sM - \lambda_1 (\sM-\sm),
\end{equation}
where $\lambda_1 \equiv (1-\exp{-\sM})/(\exp{-\sm}-\exp{-\sM})$.
Eq.~(\ref{E:upperbounds}) also implies that $\sM \ge |\sm|$, i.e., the most positive total entropy production along any trajectories is larger than the magnitude of the most negative one.
In principle, it may be possible to obtain the values of $\sM$ and $\sm$ if the underlying evolution law for the probability distribution is known. 

In a more general setting, let's imagine an oriented smooth manifold $S \subset \mathbb{R}^n$ whose surface is given by $\g(\uvec,\xvec)=0$ for $\uvec \in \mathbb{R}^n$. Suppose $\xvec:\Omega \to S$ is a random variable, and $\bar{\xvec} \equiv \avg{\xvec} = \int_\Omega \xvec \,\D\mathcal{P}(\xvec)$ is its average which again is presumably known. We would like to compute the upper bound of $\avg{f} = \int_\Omega f \,\D\mathcal{P}(\xvec)$ where $f:S \to \mathbb{R}$ is a convex function operated on $\xvec$ \emph{provided that $f(\uvec)$ is known for all $\uvec$ living on the boundary of $S$}.  In this regard, we shall proceed in two steps: (i) express any $\xvec$ inside of $S$ as a convex combination of points along boundary $\partial S$; (ii) apply Jensen's inequality and construct the upper bound for $\avg{f}$. To accomplish point (i), let's first consider another closed surface $\partial \Sdual$ [described by $\gdual(\vvec,\xvec) = 0$] where the gradient at each point on the surface corresponds to a point on $\partial S$. The form of $\gdual$ shall be prescribed later. Stokes' theorem implies that the integral of a unit normal over a closed surface is zero:
\begin{equation}\label{E:Stokes}
	0 = \oint \hat{n}\,\D a = \oint_{\partial \Sdual} \frac{\nabla \gdual(\vvec,\xvec)}{|\nabla \gdual(\vvec,\xvec)|}\,\D a(\vvec)
\end{equation}
Upon choosing 
\begin{equation}\label{E:Neumann}
	\nabla\gdual(\vvec,\xvec) = \uvec(\vvec)-\xvec,
\end{equation}
one can rewrite $\xvec$ as a convex combination of all boundary points:
\begin{equation}\label{E:convexcombo}
	\xvec = \oint_{\partial \Sdual} \uvec(\vvec) \,\D\lambda(\vvec,\xvec),
\end{equation}
where
$\D\lambda(\vvec,\xvec) \equiv \frac{1}{V}\frac{\D a(\vvec)}{|\uvec(\vvec) - \xvec|} \ge 0$ and $V \equiv \oint_{\partial \Sdual} \frac{\D a(\vvec)}{|\uvec(\vvec) - \xvec|}$.
The weight is chosen such that $\oint_{\partial \Sdual} \D \lambda(\uvec,\xvec) = 1$.

\begin{figure}[htb]
	\includegraphics[width=.45\textwidth]{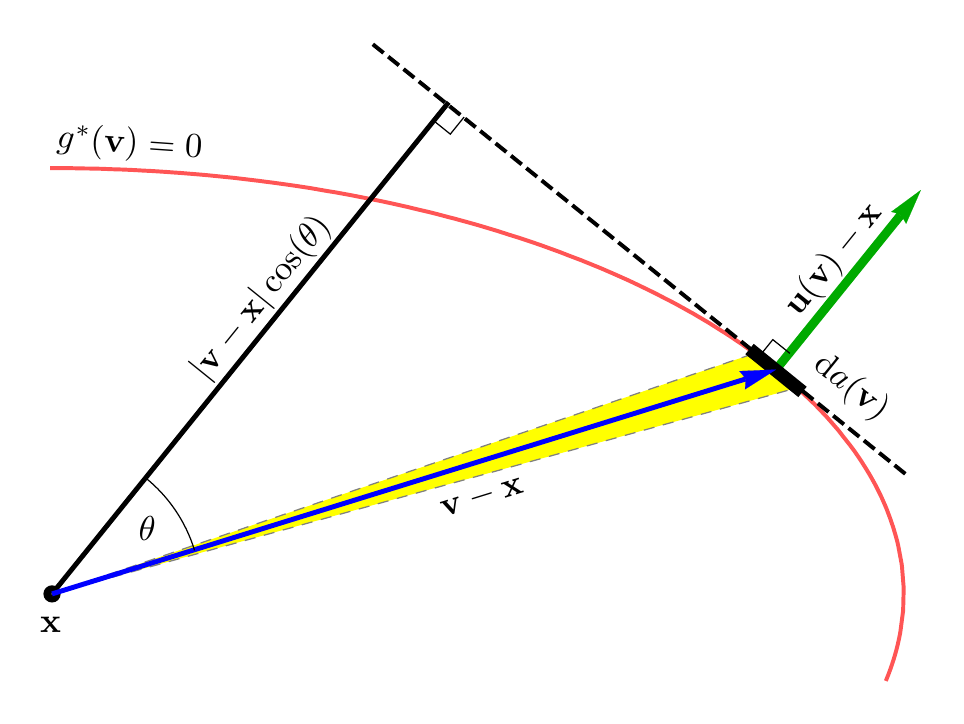}
	\caption{The surface normal at $\vvec -\xvec$ on the surface of $\Sdual$ is given by $\uvec(\vvec)-\xvec$ such that $(\uvec(\vvec)-\xvec)\cdot(\vvec-\xvec) = n$. The volume of the shaded hyperpyramid with infinitesimal base area $\D a(\vvec)$ is given by $(1/n)|\vvec-\xvec|\cos(\theta)\, \D a(\vvec) = \D a(\vvec)/|\uvec(\vvec)-\xvec|$ which gives rise to the geometrical interpretation of $V$ and $\D\lambda(\vvec,\xvec)$ in (\ref{E:convexcombo}) as the hypervolume of $\Sdual$ and the fractional volume of a hyperpyramid with base area $\D a(\vvec)$ respectively.}
	\label{fig:PolarDual}
\end{figure}

The form of $\gdual(\vvec,\xvec)$ must be so constructed that it contains $\xvec$ and Eq.~(\ref{E:Neumann}) is satisfied. The former criterion demands for all $\vvec$ on $\partial \Sdual$ that $\nabla\gdual\cdot(\vvec-\xvec) > 0$ and $\Sdual$ be convex. The positivity condition guarantees that $\nabla\gdual$ always points normally outward. This leads to
\begin{equation}\label{E:gdual}
	\gdual(\vvec,\xvec) = \sup_{\uvec\in S}\left\{(\uvec-\xvec)\cdot(\vvec-\xvec) - n   \right\},
\end{equation}
where $n$ is some positive number. Condition $\gdual(\vvec,\xvec) \le 0$ gives rise to what is known in the mathematical community as the (shifted) polar dual of $S$~\cite{AlipBord05}. A notable property of polar dual $\Sdual$ is that it is convex even though $S$ is not. This follows from the use of the supremum in (\ref{E:gdual}). It should be pointed out also that if we choose $n$ to be the dimension of the manifold, $V$ is simply the hypervolume of $\Sdual$. This is illustrated in Fig.~\ref{fig:PolarDual}.

The derivation so far does not require that $g(\uvec,\xvec)$ be differentiable everywhere. It only needs to be orientable and piecewise smooth for $\gdual(\vvec,\xvec)$ to exist, and vice versa. One of the areas where this type of convex combination is most heavily used is in geometry processing where points are represented as combinations of vertices of a control mesh (which is almost always a simplex)~\cite{JuSchaWarrDesb05}. The role of $\D\lambda(\vvec,\xvec)$ in (\ref{E:convexcombo}) is equivalent in that context to a \emph{barycentric coordinate}. When a mesh is twisted or turned, the points inside move around in space but their barycentric coordinates remain unchanged.
Suppose at point $\vcup$ on $\partial\Sdual$ that $\gdual(\vcup,\xvec)$ is not differentiable. Gradient $\nabla \gdual(\vcup,\xvec)$ in Eq.~(\ref{E:Stokes}) does not exist. The formulation is still valid if one replaces the gradient by \emph{subgradient} $\partial \gdual(\vcup,\xvec)$ defined as a set of vectors that satisfy $g(\vvec,\xvec)-g(\vcup,\xvec)\ge \partial \gdual(\vcup,\xvec) \cdot (\vvec -\vcup)$ for all $\vvec\in \Sdual$. Geometrically these vectors form a pyramid whose base represents the corresponding facet of $\partial S$ perpendicular to $\uvec(\vcup)-\xvec$. The new definition of gradient coincides with the conventional one at the points on the surface that are differentiable.  Conversely, suppose $\vvec_1-\xvec$ and $\vvec_2-\xvec$ lie on the same flat facet of $\partial\Sdual$. It is clear from (\ref{E:gdual}) that $(\vvec_2-\vvec_1)\cdot(\uvec - \xvec) = 0$. In other words, both of these vectors correspond to the same vertex $\uvec(\vvec_1)-\xvec$ on $\partial S$. Thus for any $\xvec$ inside a convex polyhedra, its convex combination is written in terms of the vertices.

Next we would like to apply Jensen's inequality to $f(\xvec)$ with $\xvec$ expressed according to (\ref{E:convexcombo}). Then the average over all possible $\xvec$ is applied. As it stands, however, the surface $\partial\Sdual$ to be integrated depends on $\xvec$ which could present a technical difficulty during the averaging process. To remedy the situation, we apply the co-area formula~\cite{Horm90}; for $h:\mathbb{R}^n \to \mathbb{R}$ such that $\nabla h$ is nowhere zero,
\begin{equation}\label{E:coarea}
	\int_{\mathbb{R}^n} F(\rvec)\,\delta(h(\rvec))\,\D^n\rvec = \oint_{h^{-1}(0)} \frac{F(\rvec)}{|\nabla h(\rvec)|}\,\D a(\rvec),
\end{equation}
where the integral on the right-hand side is over the $(n-1)$-dimensional surface defined by $h(\rvec)=0$. Eq.~(\ref{E:convexcombo}) now becomes
\begin{equation}\label{E:convexcombo2}
	\xvec = \int_{\mathbb{R}^n} \uvec(\vvec)\, w(\vvec,\xvec)\,\D^n \rvec,
\end{equation}
where
$w(\vvec,\xvec) = \delta(\gdual(\vvec,\xvec))/V(\xvec)$ and 
	$V(\xvec) = \int_{\mathbb{R}^n} \delta(\gdual(\vvec,\xvec))\,\D^n \rvec$.
In this regard, the geometrical construct is encapsulated in the weight function $w(\vvec,\xvec)$, and the integral over it gives $\int w(\rvec,\xvec)\,\D^n\rvec = 1$. An application of Jensen's inequality on $f$ immediately gives
\begin{equation}\label{E:fJensens}
	f(\xvec) \le \int_{\mathbb{R}^n} f(\uvec(\vvec))\, w(\vvec,\xvec)\,\D^n \rvec.
\end{equation}
Note that all the $\xvec$ dependence is within the weight function. Suppose we are given the value of $\bar{\xvec}$ such that
$\bar{\xvec} \equiv \avg{\xvec} = \int_{\mathbb{R}^n} \uvec(\vvec)\, \avg{w(\vvec,\xvec)}\,\D^n \rvec$.
But $\bar{\xvec}$ itself can be expanded as a convex combination 
$\bar{\xvec} = \int_{\mathbb{R}^n} \uvec(\vvec)\, w(\vvec,\bar{\xvec})\,\D^n \rvec$. Since this is true for any choice of $S$ that contains all outcomes of $\xvec$, we can therefore conclude that
\begin{equation}\label{E:wavg}
	\avg{w(\vvec,\xvec)} = w(\vvec,\bar{\xvec}). 
\end{equation}
Finally upon averaging over all $\xvec$ in (\ref{E:fJensens}) and making use of (\ref{E:wavg}) and (\ref{E:coarea}), we arrive at our first main result:
\begin{equation}\label{E:fmain}
	\avg{f(\xvec)} \le \oint_{\partial \Sdual} f(\uvec(\vvec))\,\D\lambda(\vvec,\bar{\xvec})
\end{equation}
Here the result is re-expressed as a surface integration. The equality is obtained when $\Sdual$ collapses to a point.

\section{Applications}

\subsection{Application to fluctuation relations}\label{sec:FlucRel}

\begin{figure}[htb]
	\includegraphics[width=.48\textwidth]{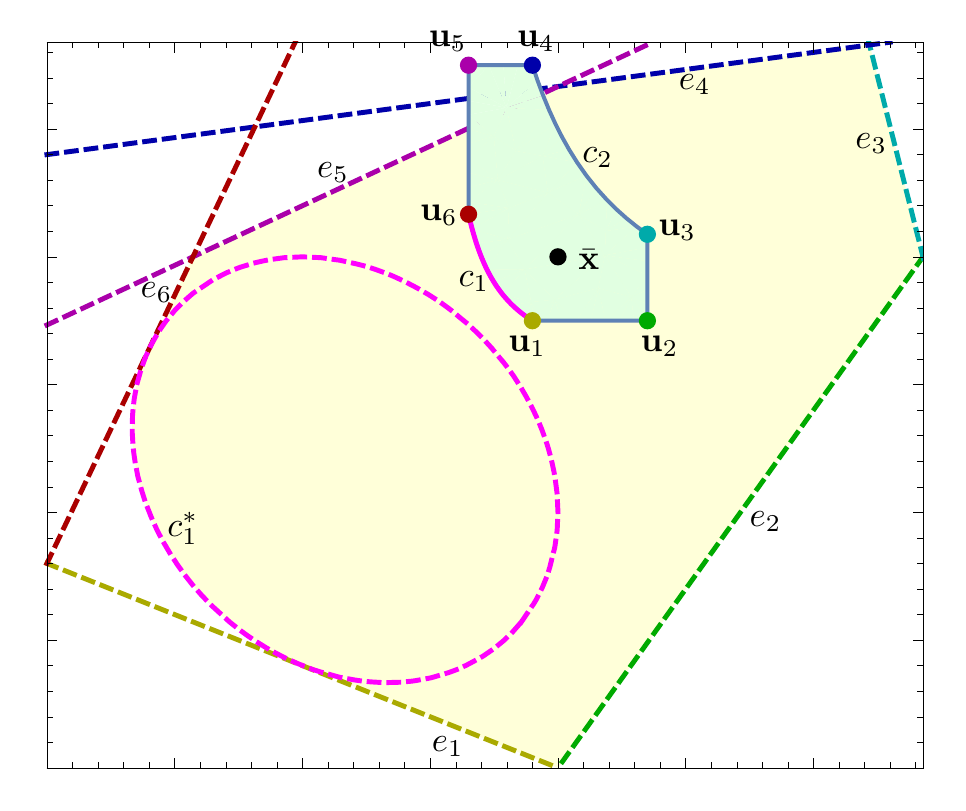}
	\caption{Possible outcomes of $\xvec$, whose average is $\bar{\xvec}$, occur inside region $S$ shown in green. Its polar dual $\Sdual$ (in yellow) lies within the boundaries formed from edges $e_1$ to $e_6$ (dual to vertices $\uvec_1$ to $\uvec_6$) and a part of curve $c_1^*$ (dual to $c_1$). Part $c_2$ of $\partial S$ between $\uvec_3$ and $\uvec_4$ is non-convex and thus has no dual correspondence on $\partial \Sdual$.}
	\label{fig:DualEntropy}
\end{figure}

Can we use inequality~(\ref{E:fmain}) to lower the upper bound of the entropy production as stated in (\ref{E:upperbounds})? 
As mentioned earlier, the trajectory-wise total entropy production can be written as the sum between the adiabatic and nonadiabatic contributions: $\ds = \dsa+\dsna$. Furthermore 
$$ \avg{\mathrm{exp}(-\ds)} = \avg{\mathrm{exp}(-\dsa)} = \avg{\mathrm{exp}(-\dsna)} = 1. $$
This means that if one treats $\xvec = (x,y) = \left(\mathrm{exp}(-\dsa), \mathrm{exp}(-\dsna)\right)$ as a pair of random variables whose average is $\bar{\xvec} = (1,1)$, one can compute the upper bound of $\avg{f}$ where $f \equiv -\ln(xy)$ given the boundaries of $\xvec$. Let's denote the maxima and minima of $\dsa$ and $\dsna$ respectively by the pairs $(\dsaM, \dsam)$ and $(\dsnaM, \dsnam)$. Possible outcomes of $\xvec$ must fall within boundaries formed from the lines joining vertices $\uvec_1$ to $\uvec_3$ and $\uvec_4$ to $\uvec_6$, and two curves $c_1$ and $c_2$ as shown in Fig.~\ref{fig:DualEntropy}. The values of $\uvec_1$ to $\uvec_6$ are given in Table~\ref{tab:uvecs}. 
Curves $c_1: \left(\mathrm{exp}(-\dsa),\mathrm{exp}(-\sM+\dsa)\right)$ and $c_2: \left(\mathrm{exp}(-\dsa),\mathrm{exp}(-\sm+\dsa)\right)$ are the boundaries that correspond to the maximum and minimum of the trajectory-wise total entropy production, respectively.
The polar dual $\Sdual$ centered at $\bar{\xvec}$ is obtained from Eq.~(\ref{E:gdual}). The result is also shown in Fig.~\ref{fig:DualEntropy}. The area is bounded by edges $e_1$ to $e_6$ and curve $c_1^*$. These are obtained from vertices $\uvec_1$ to $\uvec_6$ and curve $c_1$, respectively. (See the detailed calculations of the shape of $\Sdual$ in Appendix~\ref{sec:PolarDualConstruction}, and its area in Appendix~\ref{sec:AppB}.)

The final step of computing the upper bound for the total entropy production $\dSM$ is to perform the integral (\ref{E:fmain}) over the boundary of $\Sdual$.
The integral can be split into seven pieces along the piecewise smooth boundaries.
Each integral represents the area of a region obtained from joining $\bar{\xvec}$ to the two end points of the corresponding boundary forming either a triangle (for $e_1$ to $e_6$) or a pie shape (for $c_1^*$). 
In this problem, $f(\uvec(\vvec))$ along the outer boundary of region $i$ is simply a constant $f_i$. Eq.~(\ref{E:fmain}) thus reduces to computing the sum of $f(\uvec(\vvec))$ over all seven regions weighted by their fractional areas. We finally arrive at the new upper bound of the total entropy production:
\begin{multline}\label{E:Striangle}
	\dSM = \sum_{i=1}^7 a_i f_i 
	= \sM - (a_3+a_4)(\sM-\sm) \\
	-\big[a_2(\sM -\dsnaM -\dsam) + a_5(\sM -\dsaM -\dsnam) \big]
\end{multline}
where $a_i = A_i/\sum_{j=1}^7 A_j$ is the fractional area of region $i$. All the relevant parameters are listed in Table~\ref{tab:areas}.

It should be noted that the new upper bound $\dSM$ as given by Eq.~(\ref{E:Striangle}) is not guaranteed to be lower than that given in (\ref{E:upperbounds}) for all possible values of the parameters. By comparing the two equations, it is clear that the new upper bound will be lower if the following conditions are satisfied simultaneously: (i) $a_3+a_4 > \lambda_1$ and (ii) $a_2(\sM -\dsnaM -\dsam) + a_5(\sM -\dsaM -\dsnam) > 0$. In practice, whether these are true depends on the underlying evolution equation of the system.


\subsection{Relations between bounds on the average and the entropy production suprema for non-equilibrium steady state system}\label{s:RelnBounds}

\begin{figure}[htb]
	\includegraphics[width=.48\textwidth]{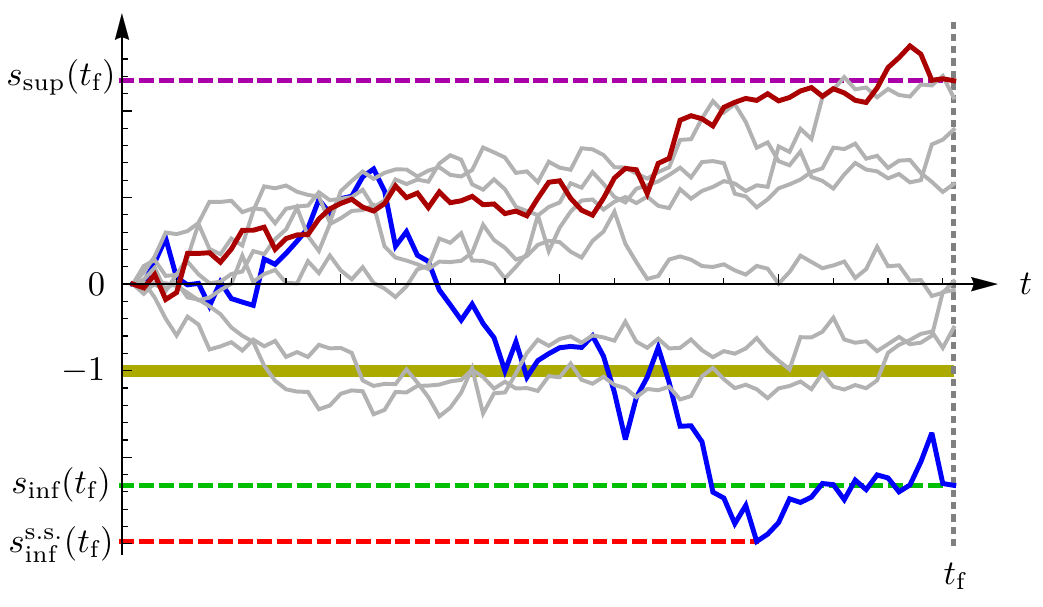}
	\caption{Entropy productions along some sampling stochastic trajectories are plotted. The supremum and infimum of the total entropy production at time $\tf$ over all trajectories are respectively represented by $\sM(\tf)$ and $\sm(\tf)$. There is only one pair of such quantities per ensemble. Each trajectory however has one entropy production infimum along its trajectory, $\smss(\tf)$, which is located at the lowest point of the trajectory during time interval $[0,\tf]$. The ensemble average over all $\smss$'s is equal to $-1$~\cite{NeriRoldJuli17}.} 
	\label{fig:trajDefn}
\end{figure}

Most recently Neri et~al.~\cite{NeriRoldJuli17} showed that, for any non-equilibrium stationary state, process $\text{exp}(-\ds)$ is a \emph{martingale}; its expected outcome of a process at time $\tf$, conditioned on a particular trajectory $\omega(0,t)$ from time $0$ up to time $t<\tf$, equals the value of the process at $t$ itself:
\begin{equation}\label{E:martingale}
	\avg{\exp{-\ds(\tf)}|\omega(0,t)} = \exp{-\ds(t)}.
\end{equation}
One may take $t=0$, where $\ds(0)=0$, in which case one gets $\avg{\exp{-\ds(\tf)}|\omega(0,t)} = 1$ for any $\omega(0,t)$. Thus the integral fluctuation relation follows naturally from the martingale property of $\exp{-\ds}$~\cite{ChetGupt11, NeriRoldJuli17}.
Employing Doob's inequality to (\ref{E:martingale}), they then obtained the cumulative distribution of the infimum of entropy production \emph{along} a trajectory during time interval $0<t<\tf$:
 \begin{equation}\label{E:infEntProd}
	\mathrm{Pr}[\smss \ge -s] \ge 1-\exp{-s},
\end{equation}
for a positive number $s$. 
The new infimum, defined by $\smss(\tf) \equiv \inf_{0\le t \le \tf} \ds(t)$, is different from the global infimum at time $\tf$ over all trajectories $\sm$ introduced at the end of Sec.~\ref{s:intro}. (Fig.~\ref{fig:trajDefn} gives an illustration of the two types of infima.)
Eq.~(\ref{E:infEntProd}) reveals that the cumulative probability distribution (left-hand side) statistically dominates over the cumulative probability distribution of an exponential random variable $s$ (right-hand side). 
Consequently, it implies that $\avg{\smss} \ge -1$. This presents a lower bound for the average of the infima. Since $\sm$ is most likely going to be much less than $-1$, it would be more advantageous if we could replace $\sm$ in (\ref{E:upperbounds}) by this number.

First we need to show that $M \equiv \exp{-\avg{\smss}} = \exp{}$ bounds $\exp{-\ds}$ from above. For the stationary state, according to (\ref{E:martingale}), let process $x$ be given by $x(\omega)\equiv \exp{-\ds(t)} = \avg{\exp{-\ds(\tf)}|\omega(0,t)}$. According to the fluctuation relation, its value is 1, which is clearly below $\exp{}$. Let $m \equiv \exp{-\sM}$ be the lower bound of $\exp{-\ds}$, where $\sM(\tf)$ is the supremum over all entropy productions at time $\tf$ as defined previously in the Introduction, and take the average in (\ref{E:IneqProof}) to be over all possible trajectories that start at time $0$ up to $t < \tf$. It is clear that $\bar{x} = \int x(\omega)\, \mathcal{D}\omega(0,t) = 1$, and for $f(x) \equiv -\ln x$, $\bar{f}(x) = \int \ds \,\mathcal{D}\omega(0,t) = \avg{\ds}$. Substituting everything into Eq.~(\ref{E:upperbounds}), we arrive at our second main result; the average entropy production is bounded according to
\begin{equation}\label{E:betterBound}
	\dS \le \dSMss = \frac{1+\exp{\sM}\left((\exp{}-1)\sM - 1 \right)}{\exp{\sM+1}-1}. 
\end{equation}
Thus for stationary stochastic processes, the bound of $\dS$ is determined solely from the trajectory-wise maximum total entropy production $\sM$. This bound is stricter than that in the case in which the system is not in a steady state. Without the martingale property, one would have to use the global infimum $\sm(\tf)$ in Eq.~(\ref{E:upperbounds}), which reduces the bound to $\sM - \exp{-|\sm|}(\sM-\sm)$ for an ensemble that contains a rare event with large negative entropy production. In such a case, the upper bound is dominated by how large $|\sm|$ is, and Eq.~(\ref{E:upperbounds}) is no longer very useful.

\begin{figure}[htb]
	\includegraphics[width=.48\textwidth]{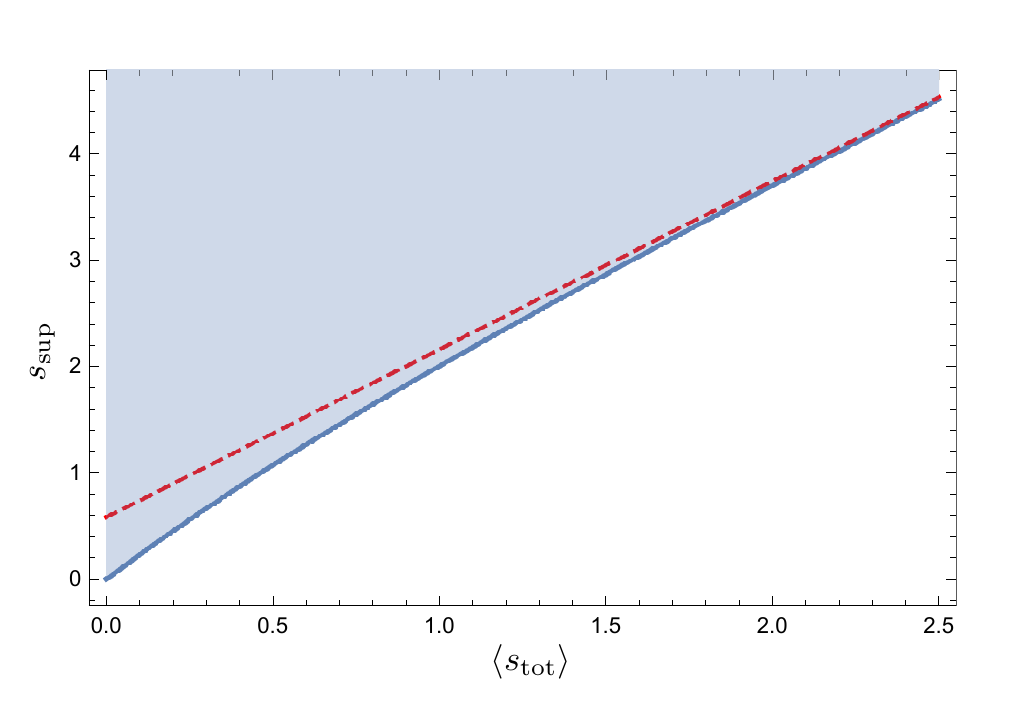}
	\caption{This plot shows the region of possible values of the average entropy production $\dS$ and the trajectory-wise total entropy production supremum $\sM$. The value of $\sM$ is bounded from below by a curve described in Eq.~(\ref{E:maxEntBound}). This bound quickly approaches $(1+\exp{}\,\dS)/(\exp{}-1)$ (dashed line) for a sufficiently large value of $\dS$.} 
	\label{fig:SvsSmax}
\end{figure}

It should be noted that Eq.~(\ref{E:betterBound}) is invertible. We can thus express the lower bound of $\sM$ if the information about $\dS$ is known. This is given by
\begin{equation}\label{E:maxEntBound}
	\sM \ge \sM^\infty + W\!\left[-\frac{\dS+1}{\exp{}-1}\,\exp{-\sM^\infty}\right],
\end{equation}
where $\sM^\infty \equiv (1+\exp{}\,\dS)/(\exp{}-1)$ is the limiting lower bound for the total entropy production supremum when $\dS$ is large, and $W[\cdot]$ denotes the Lambert-$W$ function (also known as the product logarithm function or the omega function). The Lambert-$W$ function is real when its argument is greater than $-1/\exp{}$, which is true in our case since $\dS \ge 0$ by the second law.
The relationship between the average entropy production and the lower bound of the entropy production supremum along a trajectory is shown in Fig.~\ref{fig:SvsSmax}.

At this stage, it is instructive to have a look at an example. Consider a one-dimensional overdamped system of one particle moving in a periodic potential $V(x)$ with period $L$ subjected to a constant external force $F$. This system is also known as the tilted Smoluchowski-Feynman ratchet model when $x$ is interpreted as the position of a pawl along the teeth of a ratchet in the form of $V(x)$. The model is one of a few statistical examples that can be analyzed analytically. According to this model, position $x(t)$ moves according to the following Langevin equation:
\begin{equation}\label{E:Langevin}
	\dot{x} = -\frac{V'(x)}{\gamma} + \frac{F}{\gamma} + \eta(t).
\end{equation}
Here $\gamma$ represents the friction coefficient, $\eta(t)$ is a Gaussian white noise with zero mean and autocorrelation $\avg{\eta(t)\eta(t')} = 2D \delta(t-t')$, and $D = \kB T/\gamma$ is the diffusion coefficient. 
The corresponding Fokker-Planck equation is given by $\partial_t p(x,t) = -\partial_x j(x,t)$, where $p(x,t)$ denotes the probability of finding the particle at position $x$ and time $t$, and the probability current
$$ j(x,t) = -\frac{V'(x) - F}{\gamma}\, p(x,t) - D\pdif{p(x,t)}{x}.$$

For large $t$ where the system approaches a non-equilibrium steady state, the total entropy production for a periodic potential is given by the following expression~\cite{Stra69}:
\begin{equation*}
	\ds(t) = \ln \frac{\int_{x_0}^{x_0+L} \mathrm{exp}\!\left[(V(y)-F y)/\kB T\right]\,\D y}{\int_{x(t)}^{x(t)+L} \mathrm{exp}\!\left[(V(y)-F y)/\kB T\right]\,\D y}
\end{equation*}
Gomez-Marin et al.~\cite{MariPago06} obtained an explicit form of $\ds$ in the case where $V(x)/\kB T = \ln\left[\cos(2\pi x/L) + 2 \right]$, i.e.,
\begin{equation}\label{E:R}
	\ds(t) = \frac{2\pi f}{L}(x(t)-x_0) + \ln \frac{I[x(0),f]}{I[x(t),f]},	
\end{equation}
where $f \equiv FL/(2\pi \kB T)$ is a scaled force, and $I[x,f] = f^2(\cos(2\pi x/L)+2) - f\sin(2\pi x/L) + 2$. It should be noted that the total entropy production along a trajectory depends only on the end points of the particle in this case.

\begin{figure}[htb]
	\includegraphics[width=.48\textwidth]{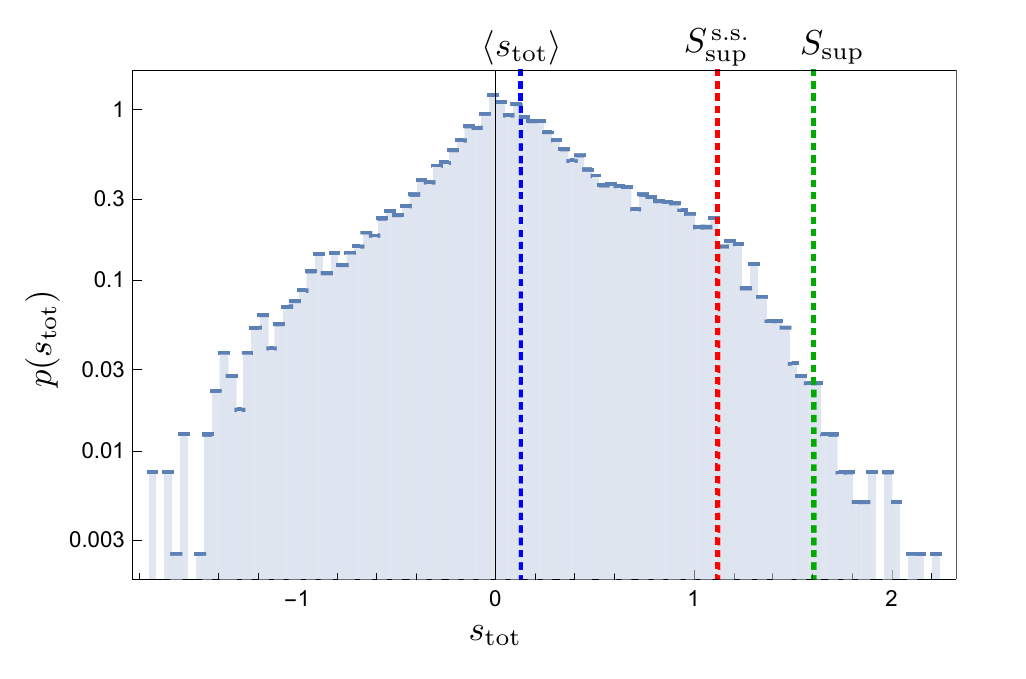}
	\caption{Probability distribution $p(\ds)$ of the trajectory-wise total entropy production $\ds$ during the time interval of $0.1\tau$ for the one-dimensional overdamped system of one particle under potential $V(x) = \kB T \ln[\cos(2\pi x/L)+2]$.  In this example, the constant applied force of $f=0.2$ acts on the particle. The histogram is generated from 10,000 independent trajectories. The average entropy production $\dS$, the steady-state maximum bound $\dSMss$ from Eq.~(\ref{E:betterBound}), and the maximum bound $\dSM$ from Eq.~(\ref{E:upperbounds}) are shown as dashed lines respectively from left to right.}
	\label{fig:RatchetGraph}
\end{figure}

To obtain an upper bound for $\dS$, we numerically integrate Eq.~(\ref{E:Langevin}) from time 0 to $100 \tau$, where $\tau \equiv L^2/D$ is the natural time scale of the system. We however take the starting position to be at $x_0=x(99.9\tau)$ to ensure that the system reasonably approaches a steady state prior to collecting the statistics. The force is taken to be $f=0.2$. Fig.~\ref{fig:RatchetGraph} shows the distribution of the entropy productions from 10,000 trajectories. The average is denoted by the dashed line on the left, while the predicted upper bound $\dSMss$ from Eq.~(\ref{E:betterBound}) is in the middle, and $\dSM$ from Eq.~(\ref{E:upperbounds}) is on the right. 
It is clear that $\dSM$ overestimates the actual mean appreciably even for the time interval as small as $0.1\tau$ due to the dominance of $|\sm|$ in (\ref{E:upperbounds}), while $\dSMss$ is unaffected by it.

\section{Remarks and conclusions}\label{sec:conclusion}
We would like to end this paper with a few remarks. At the end of the previous section, we obtained a stronger upper bound for the average entropy production based only on the supremum value of the entropy production at a final time (and conversely a lower bound for the entropy production supremum based on the average value) for a system in a non-equilibrium stationary state. Our analysis is based on the fact that the average of the infima of entropy productions is not less than $-1$~\cite{NeriRoldJuli17}. Our results for the upper bound of $\dS$ and the lower bound of $\sM$ are valid \emph{on average}. In other words, without enough statistics, there might exist an ensemble where these aren't true---in the same sense that the fluctuation relation $\avg{\exp{-\ds}} = 1$ doesn't always hold for a small sample size. (For example, see Fig. 2 in Ref.~\cite{MariPago06}.) It remains to be investigated whether one could obtain a more precise statement about the upper bound of $\dS$ (or the lower bound of $\sM$) in the spirit of Eq.~(\ref{E:infEntProd}) using inequality~(\ref{E:upperbounds}) as a constraint.

In principle one can employ inequality (\ref{E:IneqProof}) to find, not only the upper bound of a convex function $f$, but also its variance: $\text{Var}(f) = \avg{f^2} - \avg{f}^2$. This occurs when $\avg{f^2}$ is maximized while $\avg{f}^2$ is minimized. The former can be bounded with the present technique since $f^2$ is also convex. The latter happens when $\avg{f} = f(\bar{x})$. Thus,
\begin{equation*}
\begin{split}
	\mathrm{Var}(f) &\le \lambda_{\bar{x}}(f^2(M)-f^2(m)) + f^2(m) - f^2(\bar{x}) \\
	&= \bar{f}_\text{sup}(f(M)+f(m)) - f(m)f(M) - f^2(\bar{x}).
\end{split}
\end{equation*}
In our case, the variance of the total entropy productions along all stochastic trajectories is bounded by
\begin{equation}\label{E:Svar}
	\mathrm{Var}(\dS) \le (\sM + \sm)\Smax - \sM\sm.
\end{equation}
Extending the result to a larger parameter space is a laborious exercise in algebra. 
Recently there has been a great deal of interest in examining the uncertainty relation for some stochastic current $J$, where the entropy production rate is only one of the examples, in terms of the steady-state dissipation rate $\sigma_\star$~\cite{PietBaraSeif16,GingHoroPeruEngl16,PoleLazaEspo16}: $\mathrm{Var}(J)/\avg{J}^2 \ge 2/\sigma_\star$. The analysis that leads to (\ref{E:Svar}) could be useful in providing the opposite bound. This shall be explored further in a future work.

In conclusion, by expressing a random variable within a domain as a convex combination of the boundary points, it is possible to apply Jensen's inequality to a convex function over the variable. If the average of this random variable is known, then the average of the upper bound of the convex function will be known as well. We applied this result to obtain the upper bound of the total entropy production based on the extrema of trajectory-wise total entropy production, and both adiabatic and nonadiabatic entropy productions. In the case of a non-equilibrium stationary state, the upper bound only depends on the entropy production supremum. Conversely, the entropy production supremum can also be bounded from below given the average entropy production.

\begin{table}[hbt]
  \caption{\label{tab:uvecs}Coordinates of the boundary points $\uvec_1$ to $\uvec_6$ as shown in Fig.~\ref{fig:DualEntropy}.}
  \begin{ruledtabular}
  \begin{tabular}{c c c c}
    \phantom{X}$\uvec_1$ & \phantom{X}$\left(\exp{-\sM+\dsnaM}, \exp{-\dsnaM} \right)$\phantom{X} &
    \phantom{X}$\uvec_4$ & \phantom{X}$\left(\exp{-\sm+\dsnam}, \exp{-\dsnam} \right)$ \phantom{\bigg|} \\[8pt]
    \phantom{X}$\uvec_2$ & $\left(\exp{-\dsam}, \exp{-\dsnaM} \right)$ &
    \phantom{X}$\uvec_5$ & $\left(\exp{-\dsaM}, \exp{-\dsnam} \right)$ \\[8pt]
    \phantom{X}$\uvec_3$ & $\left(\exp{-\dsam}, \exp{-\sm+\dsam} \right)$ &   
    \phantom{X}$\uvec_6$ & $\left(\exp{-\dsaM}, \exp{-\sM+\dsaM} \right)$ \\[8pt]
  \end{tabular}
  \end{ruledtabular}
\end{table}

\begin{table}[hbt]
  \caption{\label{tab:areas}Areas of various regions in $\Sdual$ formed from joining $\bar{\xvec}$ and the two end points of each boundary. Below function 
$F[t] \equiv$ $\frac{2\,\exp{-\sM}}{1-\exp{-\sM}}\bigg[\kappa 
  \tan^{-1}\!\big[\kappa(t - \exp{-\sM}) \big]
  - \frac{1-t}{t^2 + \exp{-\sM}(1 - 2t)} \bigg]$ where $\kappa \equiv \big(\exp{-\sM}(1-\exp{-\sM})\big)^{-1/2}$
  }
  \begin{ruledtabular}
  \begin{tabular}{ c  c  c }
    bdry &  \hspace{.9in} Area $A_i$ \hspace{.9in} & $f_i$  \\
    \hline 
    $e_1$ & $\frac{1}{(1-\exp{-\dsnaM})(1+\exp{-\sM + 2\dsnaM}(1-2\,\exp{-\dsnaM} ))}$ & $\sM$ \\[3pt]
    $e_2$ & $\frac{1}{(\exp{-\dsam}-1)(1-\exp{-\dsnaM})}$ & $\dsam + \dsnaM$ \\[3pt]
    $e_3$ & $\frac{1}{(\exp{-\dsam}-1)(\exp{-\sm+\dsam}-1 + \exp{-\dsnam}(1-\exp{\dsam}))} $ & $\sm$ \\[3pt]
    $e_4$ & $\frac{1}{(\exp{-\dsnam}-1)(\exp{-\sm+\dsnam}-1 + \exp{-\dsam}(1-\exp{\dsnam}))} $ & $\sm$ \\[3pt]
    $e_5$ & $\frac{1}{(\exp{-\dsnam}-1)(1-\exp{-\dsaM})} $ & $\dsaM+\dsnam$ \\[3pt]
    $e_6$ & $\frac{1}{(1-\exp{-\dsaM})(1+\exp{-\sM + 2\dsaM}(1-2\,\exp{-\dsaM} ) )}$ & $\sM$ \\[3pt]
    $c_1^*$ & $F\big[\exp{-\sM+\dsnaM} \big] - F\big[ \exp{-\dsaM} \big]$ & $\sM$ \\[3pt]
  \end{tabular}
  \end{ruledtabular}
\end{table}

\newcommand{\xb}{\bar{x}}
\newcommand{\yb}{\bar{y}}

\appendix

\newcommand{\W}[2]{W_{ #1 \shortleftarrow #2 }}
\newcommand{\Wmat}{\mat{W}}
\newcommand{\dsK}{\sig_{K}}

\section{A general trajectory-wise entropy for an inhomogeneous Markov jump process}\label{sec:entropyIsBounded}
In the main text, we presented the abstract definition of a trajectory-wise entropy as a Randon-Nikod\'ym derivative between two measures. Here we would like to give an explicit formulation of this type of entropy for a particular class of problem---an inhomogeneous Markov jump process.
Consider a system with a finite number of states. We are interested in the system whose $P_i(t)$, the probability that the system is in state $i \in S$ at time $t$, obeys an inhomogeneous master equation:
\begin{equation}\label{E:MasterEqn}
	\dif{P_i}{t} = \sum_{j} \W{i}{j}(t) P_j(t)	
\end{equation}
Here $\W{i}{j}(t)$ describes the transition rate from state $j$ to state $i$ at time $t$ of an inhomogeneous Markov process $X = \{X(t): t_0 \le t \le \tf \}$. Due to the conservation of total probabilities among all states,
	$\sum_i \W{i}{j}(t) = 0.$
Consider the case in which $\Wmat(t)$ satisfies the condition that $\W{i}{j}(t) > 0$ for some $t$ iff $\W{j}{i}(t) > 0$ for all $t$. In other words, the transition rate from state $j$ to $i$ is nonzero if and only if the rate from $i$ to $j$ is also nonzero.
Let $n_t$ denote the number of times that $X$ jumps between $[t_0,\tf]$. These jumps occur at times 
$t_1 = \inf\{t>t_0: X(t) \ne X(t_0)\}$, \ldots, $t_k = \inf\{t> t_{k-1}: X(t) \ne X(t_{k-1})\}$, \ldots, $t_n$,
and they take the system through states $i_0, \ldots, i_n \in S$ satisfying $i_k \ne i_{k+1}$. A trajectory in orbit space $\Omega$ is then defined by
\begin{multline*}
	\Atraj = \{\omega \in \Omega: n_t(\omega) = n, X(t_0) = i_0, \\ X(t_k(\omega)) = i_k, k = 1,\ldots, n\}.
\end{multline*}
For a sufficiently small $\delta t_k$ around $t_k$, we claim that the probability that trajectory $\Atraj$ be in $\Afam$, where
\begin{multline*}
	\Afam = \{ \omega \in \Omega: n_t(\omega) = n, \\ X(t_0) = i_0, 
	X(t'_k(\omega)) = i_k, \\
	t_k-\delta t_k < t'_k < t_k + \delta t_k, k = 1,\ldots, n\},
\end{multline*}
is given by
\begin{multline}\label{E:PF}
	\Pforward{[t_0,\tf]}{\Afam} = \int_{t_1-\delta t_1}^{t_1+\delta t_1}\D s_1 \cdots \int_{t_n-\delta t_n}^{t_n+\delta t_n}\D s_n \\
	\W{i_n}{i_{n-1}}(s_n) \cdots \W{i_1}{i_0}(s_1) P_{i_0}(t_0) \\
	\mathrm{exp}\!\left[-\sum_{k=0}^n \int_{s_k}^{s_{k+1}} r_{i_k}(u)\,\D u \right].
\end{multline}
Here $r_{i_k}(u) = \sum_{j\ne i_k} \W{j}{i_k}(u)$ is the rate of exiting state $i_k$ at time $u$, $s_0 = t_0$, and $s_{n+1} = \tf$.

Consider another Markov process $\{X^\dagger(u): t_0\le u \le \tf\}$ generated by $\Kmat$ which is a time and/or state dependent function of $\Wmat$. The master equation that describes this process would have the same form as (\ref{E:MasterEqn}). The probability distribution of a trajectory visiting a set of states $\{i_0^\dagger,\ldots,i_n^\dagger\}$ by undergoing transitions close to times $\{t_1,\ldots,t_n\}$ is given by
\begin{multline}\label{E:PZ}
	\PKforward{[t_0,\tf]}{\Kfam} = \int_{t_1-\delta t_1}^{t_1+\delta t_1}\D s_1 \cdots \int_{t_n-\delta t_n}^{t_n+\delta t_n}\D s_n \\
	\times\K{i^\dagger_n}{i^\dagger_{n-1}}(s_n) \cdots \K{i^\dagger_1}{i^\dagger_0}(s_1) P_{i^\dagger_0}(t_0) \\
	\times\mathrm{exp}\!\left[-\sum_{k=0}^n \int_{s_k}^{s_{k+1}} \rK_{i^\dagger_k}(u)\,\D u \right].
\end{multline}
The sequence of states is so chosen to match those in (\ref{E:PF}). This implies by construction that $\Pforward{[t_0,\tf]}{\Afam}$ and $\PKforward{[t_0,\tf]}{\Kfam}$ are continuous with respect to one another, i.e., $\Pforward{[t_0,\tf]}{\Afam} = 0$ iff $\PKforward{[t_0,\tf]}{\Kfam} = 0$. Therefore the Radon--Nikod\'ym derivative exists and can be read off from (\ref{E:PF}) and (\ref{E:PZ}):
\begin{multline}\label{E:dPdPK}
	\frac{\D\Pforward{[t_0,\tf]}{\Atraj}}{\D\PKforward{[t_0,\tf]}{\Ktraj}} = \frac{P_{i_0}(t_0)}{P_{i^\dagger_0}(t_0)}\prod_{k=0}^{n-1} \frac{\W{i_{k+1}}{i_k}(t_{k+1})}{\K{i^\dagger_{k+1}}{i^\dagger_k}(t_{k+1})} \\
	\times \frac{\mathrm{exp}\!\left[-\sum_{k=0}^n \int_{t_k}^{t_{k+1}} r_{i_k}(u)\,\D u\right]}{\mathrm{exp}\!\left[-\sum_{k=0}^n \int_{t_k}^{t_{k+1}} \rK_{i^\dagger_k}(u)\,\D u\right]} ,
\end{multline}
where $t_{n+1} = \tf$. We thus define the trajectory-dependent change in entropy of the forward process with respect to process $K$ as
\begin{equation}\label{E:entropydef}
	\dsK(\Atraj) \equiv \ln\!\left[\frac{\D\Pforward{[t_0,\tf]}{\Atraj}}{\D\PKforward{[t_0,\tf]}{\Ktraj}}\right].
\end{equation}
To see that $\dsK(\Atraj)$ is bounded, it is enough to show that $|\dsK(\Atraj)| < \infty$. One must realize that, by construction, $\K{i^\dagger_{k+1}}{i^\dagger_k}(t_{k+1}) > 0$ as long as $\W{i_{k+1}}{i_k}(t_{k+1}) > 0$. Together with the fact that $P_{i_0}(t_0) > 0$ and $P_{i^\dagger_0}(t_0) > 0$, we can conclude that every term on the right hand side of (\ref{E:dPdPK}) is bounded, and so must $|\dsK(\Atraj)|$ be. Therefore there must exist a supremum and an infimum for $\dsK(\Atraj)$.

As an illustration, consider a backward Markov process $\{X(t_0+\tf -u): t_0 \le u \le \tf\}$ where states are visited in the backward order. The protocol for creating this process is also obtained from the forward one by reversing the time, i.e.,
\begin{equation*}
	\WB{i}{j}(t) = \W{i}{j}(t_0+\tf -t), \quad \rB_i(t) = r_i(t_0+\tf - t).
\end{equation*}
Combining both transformations leads to the following identifications:
\begin{equation}\label{E:KtoWB}
\begin{split}
	\K{i^\dagger_{k+1}}{i^\dagger_k}(t_{k+1}) &= \WB{i^\text{B}_{k+1}}{i^\text{B}_k}(t_0+\tf-t_{n-k}) \\
	&= \W{i_{n-k-1}}{i_{n-k}}(t_{n-k}),
\end{split}
\end{equation}
and
\begin{equation}\label{E:rKtorB}
\begin{split}
	-\sum_{k=0}^n \int_{t_k}^{t_{k+1}} &\rK_{i^\dagger_k}(u)\,\D u \\
	&= -\sum_{k=0}^n \int_{t_0+\tf-t_{n-k+1}}^{t_0+\tf+t_{n-k}} \rB_{i_{n-k}}(u)\,\D u  \\
	&= -\sum_{k=0}^n \int_{t_0+\tf-t_{n-k+1}}^{t_0+\tf+t_{n-k}} r_{i_{n-k}}(t_0+\tf-u)\,\D u  \\
	&= -\sum_{k=0}^n \int_{t_k}^{t_{k+1}} r_{i_k}(u) \,\D u.
\end{split}
\end{equation}
The last equality involves a change of variables and reindexing the summation.
Substituting both (\ref{E:KtoWB}) and (\ref{E:rKtorB}) into (\ref{E:dPdPK}), while choosing $P_{i^\dagger_0}(t_0) = P_{i^\text{B}_0}(t_0) = P_{i_n}(\tf)$, results in
\begin{equation}\label{E:DStotal}
	\exp{\sig_\text{tot}(\Atraj)} \equiv \frac{\D\Pforward{[t_0,\tf]}{\Atraj}}{\D\Pbackward{[t_0,\tf]}{\Btraj}} = \frac{P_{i_0}(t_0)}{P_{i_n}(\tf)}\prod_{k=0}^{n-1} \frac{\W{i_{k+1}}{i_k}(t_{k+1})}{\W{i_{k}}{i_{k+1}}(t_{k+1})}.
\end{equation}
Notice that the expected waiting time factors exactly cancel one another.

The reason we can identify $\sig_\text{tot}(\Atraj)$ as the total entropy production along trajectory $\Atraj$ is most conveniently illustrated with a thermodynamic example. Eq.~(\ref{E:DStotal}) consists of (i) the boundary term involving probabilities of the initial and the final states, and (ii) the process term. The boundary term can be identified with $\Delta\sig_\text{sys}(\Atraj)$ if the system's trajectory-wise entropy at time $t$ is the (negative) logarithm of the probability to find the system in that particular state at time $t$: $s_\text{sys}(\Atraj,t) \equiv -\ln P_{i(t)}(t)$.
The process term requires a little more work.
In the case where detailed balance exists,
$$ \W{i_{k+1}}{i_k}(t_{k+1})\, \pi_{i_k}(t_{k+1}) = \W{i_k}{i_{k+1}}(t_{k+1}) \, \pi_{i_{k+1}}(t_{k+1}), $$
where $\pi_{i_k}(t)$ denotes the time-dependent stationary (or equilibrium) distribution of state $i_k$ at time $t$ of a canonical ensemble. If we assume that the system is in thermal contact with a reservoir whose inverse temperature is $\beta$, then the stationary distribution $\pi_{i_k}(t)$ is merely a Boltzmann factor:
\begin{equation*}
	\pi_{i_k}(t) = \frac{\exp{-\beta E_{i_k}(t)}}{Z(t)}
\end{equation*}
Here $E_{i_k}(t)$ is the energy of state $i_k$ at time $t$, and $Z(t)$ is the partition function at that time. The process term of (\ref{E:DStotal}) becomes
\begin{equation*}
	\beta \sum_{k=0}^{n-1} \left\{E_{i_k}(t_{k+1}) - E_{i_{k+1}}(t_{k+1})\right\} = \beta \sum_{k=0}^{n-1} \delta Q(t_{k+1}),
\end{equation*}
where $\delta Q(t_{k+1})$ is the heat \emph{produced by the reservoir} (or equivalently heat dissipation from the system) at time $t_{k+1}$ resulting from the energy difference between states $i_k$ and $i_{k+1}$. The process term therefore reflects the change in entropy of the environment due to the production of heat by the reservoir:
\begin{equation}\label{E:DSenv}
	\sig_\text{env}(\Atraj) = \beta Q = \beta \sum_{k=0}^{n-1} \delta Q(t_{k+1})
\end{equation}
The process term was introduced for the first time by Lebowitz and Spohn as an \emph{action functional} in their discussion of Gallavotti--Cohen-like symmetry of the generating function of $\sig_\text{env}$~\cite{LeboSpoh98}. 

Finally, with the above identifications, Eq.~(\ref{E:DStotal}) becomes
\begin{equation}\label{E:EntropySplit}
	\sig_\text{tot}(\Atraj) = \Delta\sig_\text{sys}(\Atraj) + \sig_\text{env}(\Atraj).
\end{equation}
For a more general case in which detailed balance does not occur or there is no thermodynamics connection, one can still take Eq.~(\ref{E:DStotal}) as a defining expression for total entropy production along a trajectory.


\section{Constructing the polar dual to the domain of possible entropy values}\label{sec:PolarDualConstruction}

We would like to compute the polar dual $\Sdual$ to the domain $S$ of possible outcomes of a random variable $\xvec = \left(\mathrm{exp}(\dsa),\mathrm{exp}(\dsna) \right)$ bounded by lines joining vertices $\uvec_1$ to $\uvec_3$ and vertices $\uvec_4$ to $\uvec_6$. Curves $c_1$ and $c_2$ link vertices $\uvec_1$ and $\uvec_6$, and $\uvec_3$ and $\uvec_4$ respectively. The values of $\uvec_1$ to $\uvec_6$ and the descriptions of $c_1$ and $c_2$ are given in the main text.
As described in the main text, the boundary of the polar dual $\Sdual$ of a domain $S$ can be computed from $\gdual(\vvec,\bar{\xvec}) = 0$ where
\begin{equation}\label{E:gdual2}
	\gdual(\vvec,\bar{\xvec}) = \sup_{\uvec\in S}\left\{(\uvec-\bar{\xvec})\cdot(\vvec-\bar{\xvec}) - n   \right\},
\end{equation}
for a positive integer $n$ chosen to be the dimension of $S$ ($n=2$ in this case). Here we anchor the polar dual about $\bar{\xvec} = (\xb,\yb) = (1,1)$, whose values are obtained from the fluctuation theorems. It is possible to construct $\gdual$ by computing different parts separately, then subsequently piecing them together. In this vein, we first calculate the dual of a vertex $\uvec_i$, then that of curve $c_1$.
The first part is simple to compute.
Suppose $\uvec_i$ is such that it yields the supremum of the right-hand side of (\ref{E:gdual2}). Condition $\gdual = 0$ in this case leads to
\begin{equation}\label{E:dualpoint}
	(u_{ix}-\xb)(v_x-\xb)+(u_{iy}-\yb)(v_y-\yb) = n,
\end{equation}
which is simply a straight line equation in $\vvec\in \Sdual$.

\begin{figure}[htb]
	\includegraphics[width=.48\textwidth]{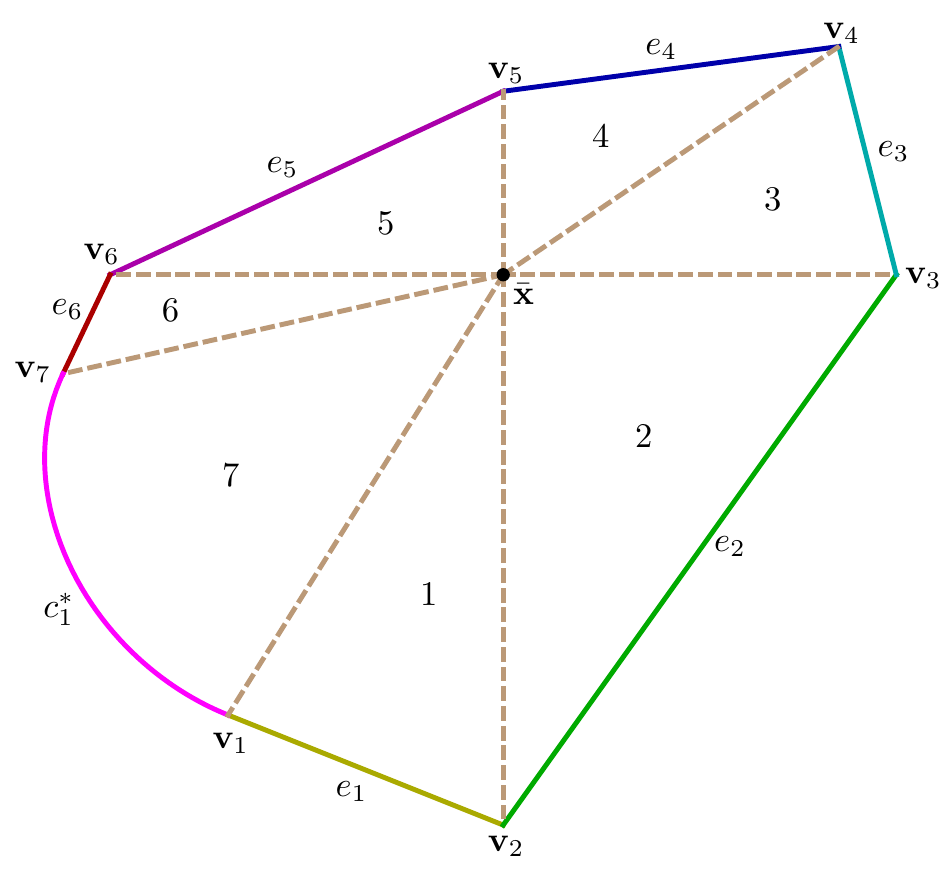}
	\caption{The area of the polar dual $\Sdual$ to domain $S$ is divided into seven regions.}
	\label{fig:DualZones}
\end{figure}

The second part involves more work. It is convenient to shift the origin to $\bar{\xvec}$, then curve $c_1$ is expressible as $\uvec = \left(u_x, -\yb + \alpha/(u_x+\xb) \right)$ for $\mathrm{exp}(-\dsaM) \le u_x + \xb \le \alpha\,\mathrm{exp}(\dsnaM)$ and $\alpha \equiv \mathrm{exp}(-\sM)$. Then (\ref{E:gdual2}) becomes 
$$\gdual = \sup_{u_x}\{G(u_x)\},$$
where 
$$	G(u_x) \equiv u_x v_x + \left(-\yb+\frac{\alpha}{u_x+\xb}\right)v_y-n.
$$
The extrema of $G(u_x)$ occur at $u_x+\xb = \pm \sqrt{\alpha \,v_y/v_x}$. The second derivative test reveals that the positive term gives the maximum, and requires that both $v_x$ and $v_y$ are negative. The condition $\gdual = 0$ becomes
$ 2\sqrt{\alpha\, v_x v_y} = -(n + \xb\, v_x + \yb\, v_y)$.
The unshifted expression is therefore
\begin{equation}\label{E:dualcurve}
	4\alpha (v_x-\xb)(v_y-\yb) = \left(n+\xb(v_x-\xb)+\yb(v_y-\yb)\right)^2.
\end{equation}
Geometrically, Eq.~(\ref{E:dualcurve}) is an equation for a scaled ellipse whose principal axes rotated by $45^\circ$.
Fig.~\ref{fig:DualZones} shows the final result after combining (\ref{E:dualpoint}) and (\ref{E:dualcurve}). The coordinates of all the controlled vertices are listed in Table~\ref{tab:vvecs}.

\section{Computing the upper bound of the total entropy production}\label{sec:AppB}
In the main text, we established that the upper bound of the average of a convex function $f$ over $S$ can be calculated from the coordinates of the domain's boundary according to
\begin{equation}\label{E:main}
	\avg{f(\xvec)} \le \oint_{\partial \Sdual} f(\uvec(\vvec))\,\D\lambda(\vvec,\bar{\xvec}),
\end{equation}
where, in the present two dimensional context,
\begin{equation*}
	\D\lambda(\vvec,\xvec) \equiv \frac{1}{A}\frac{\D s(\vvec)}{|\uvec(\vvec) - \xvec|} \ge 0 \quad \& \quad
	A \equiv \oint_{\partial \Sdual} \frac{\D s(\vvec)}{|\uvec(\vvec) - \xvec|}.
\end{equation*}
Here $\D s(\vvec)$ represents the infinitesimal length element along $\partial\Sdual$ parametrized by $\vvec$.
To apply this formula, one first computes area $A$ of the polar dual. As shown in Fig.~\ref{fig:DualZones}, the area is divided into seven regions, most of which are made up of triangles. These areas can therefore be computed quite easily from the three corner vertices without having to evaluate any integrals. Suppose a triangle is made up from joining vertices $\vec{a}$, $\vec{b}$ and $\bar{\xvec}$. Its area is given by
\begin{equation}\label{E:TriArea}
	A_\triangle = \frac{1}{2}\left|(\vec{a}-\bar{\xvec})\times (\vec{b}-\bar{\xvec})\right|.
\end{equation}

The calculation of the area of region 7 is more involved. First one needs to parametrize $c_1^*$ and $u(\vvec)$ so that $\int 1/|\uvec(\vvec)-\xvec|\,\D s(\vvec)$ can be computed. There are many choices of parametrization. The simplest one is given by
\begin{equation}
\begin{split}
	\uvec(t) &= \left(t, \exp{-\sM}/t \right), \\
	\vvec(t) &= \bar{\xvec} - \frac{n}{\yb\,t^2 + \exp{-\sM}(\xb - 2t)}\left(\exp{-\sM}, t^2 \right),
\end{split}
\end{equation}
for $\exp{-\dsaM} \le t \le \exp{-\sM+\dsnaM}$. 
The integral can be written as
\begin{equation*}
\begin{split}
	A_7 &= \int_{\exp{-\dsaM}}^{\exp{-\sM+\dsnaM}} \frac{1}{|\uvec(t)-\bar{\xvec}|}\,\left|\frac{\D\vvec}{\D t} \right|\,\D t \\
	&= 2 n\, \exp{-\sM}\int_{\exp{-\dsaM}}^{\exp{-\sM+\dsnaM}} \frac{t}{\left(\yb\, t^2 + \exp{-\sM}(\xb - 2t) \right)^2}\, \D t \\
	&= F\big[\exp{-\sM+\dsnaM} \big] - F\big[ \exp{-\dsaM} \big],
\end{split}
\end{equation*}
where, after substituting $\xb = \yb = 1$ and $n=2$,
\begin{multline*}\label{E:V7}
	F[t] \equiv \frac{2\,\exp{-\sM}}{1-\exp{-\sM}}\bigg[\frac{1}{\sqrt{\exp{-\sM}(1-\exp{-\sM})}} \\
	\tan^{-1}\!\left(\frac{t-\exp{-\sM}}{\sqrt{\exp{-\sM}(1-\exp{-\sM})}} \right) -\frac{1-t}{t^2 + \exp{-\sM}(1 - 2t)}  \bigg].
\end{multline*}
The values of all the areas are summarized in Table~II of the main text.

\widetext
\renewcommand{\arraystretch}{2.5}
\begin{center}
\begin{table}[hbt]
  \caption{\label{tab:vvecs}Coordinates of the boundary points $\vvec_1$ to $\vvec_7$ of the polar dual $\Sdual$ as shown in Fig.~\ref{fig:DualZones}.}
  \begin{tabular}{c  c  c}
  	\hline \hline
    \hspace{10pt}$\vvec_i$ \hspace{10pt} & \hspace{1in} $\dfrac{v_{ix}-\xb}{n}$ \hspace{1in} & \hspace{1in} $\dfrac{v_{iy}-\yb}{n}$ \hspace{1in}   \\[5pt]
    \hline 
    $\vvec_1$ & 
    $-\dfrac{1}{\xb + \exp{-\sM+2\dsnaM}\left(\yb - 2\,\exp{-\dsnaM} \right)}$ &
     $-\dfrac{1}{\yb + \exp{\sM-2\dsnaM}\left(\xb - 2\,\exp{-\sM+\dsnaM} \right) }$  \\
    $\vvec_2$ & $0$ & $-\dfrac{1}{\yb - \exp{-\dsnaM}}$ \\
    $\vvec_3$ & $\dfrac{1}{\exp{-\dsam} - \xb}$ & $0$ \\
    $\vvec_4$ & 
    $\dfrac{\exp{-\dsnam}}{\exp{-\dsam}\left(\exp{-\dsnam}-\yb \right) - \left(\xb\,\exp{-\dsnam} -\exp{-\sm}\right) }$ &
    $\dfrac{\exp{-\dsam}}{\exp{-\dsam}\left(\exp{-\dsnam}-\yb \right) - \left(\xb\,\exp{-\dsnam} -\exp{-\sm}\right) }$ \\
    $\vvec_5$ & $0$ & $\dfrac{1}{\exp{-\dsnam}-\yb}$ \\
    $\vvec_6$ & $-\dfrac{1}{\xb - \exp{-\dsaM}}$ & $0$ \\
    $\vvec_7$ & 
    $-\dfrac{1}{\xb + \exp{\sM-2\dsaM}\left(\yb - 2\,\exp{-\sM + \dsaM} \right)}$ &
    $-\dfrac{1}{\yb + \exp{-\sM+2\dsaM}\left(\xb - 2\,\exp{-\dsaM} \right)}$ \\[10pt]
   \hline   \hline
  \end{tabular}
\end{table}
\end{center}

\begin{acknowledgments}
The author is grateful to Udomsilp Pinsook for his insightful input and valuable suggestions. Funding from the Development and Promotion of Science and Technology Talents Project (DPST) Grant No. 028/2555 is acknowledged.
\end{acknowledgments}


\begin{thebibliography}{30}%
\makeatletter
\providecommand \@ifxundefined [1]{%
 \@ifx{#1\undefined}
}%
\providecommand \@ifnum [1]{%
 \ifnum #1\expandafter \@firstoftwo
 \else \expandafter \@secondoftwo
 \fi
}%
\providecommand \@ifx [1]{%
 \ifx #1\expandafter \@firstoftwo
 \else \expandafter \@secondoftwo
 \fi
}%
\providecommand \natexlab [1]{#1}%
\providecommand \enquote  [1]{``#1''}%
\providecommand \bibnamefont  [1]{#1}%
\providecommand \bibfnamefont [1]{#1}%
\providecommand \citenamefont [1]{#1}%
\providecommand \href@noop [0]{\@secondoftwo}%
\providecommand \href [0]{\begingroup \@sanitize@url \@href}%
\providecommand \@href[1]{\@@startlink{#1}\@@href}%
\providecommand \@@href[1]{\endgroup#1\@@endlink}%
\providecommand \@sanitize@url [0]{\catcode `\\12\catcode `\$12\catcode
  `\&12\catcode `\#12\catcode `\^12\catcode `\_12\catcode `\%12\relax}%
\providecommand \@@startlink[1]{}%
\providecommand \@@endlink[0]{}%
\providecommand \url  [0]{\begingroup\@sanitize@url \@url }%
\providecommand \@url [1]{\endgroup\@href {#1}{\urlprefix }}%
\providecommand \urlprefix  [0]{URL }%
\providecommand \Eprint [0]{\href }%
\providecommand \doibase [0]{http://dx.doi.org/}%
\providecommand \selectlanguage [0]{\@gobble}%
\providecommand \bibinfo  [0]{\@secondoftwo}%
\providecommand \bibfield  [0]{\@secondoftwo}%
\providecommand \translation [1]{[#1]}%
\providecommand \BibitemOpen [0]{}%
\providecommand \bibitemStop [0]{}%
\providecommand \bibitemNoStop [0]{.\EOS\space}%
\providecommand \EOS [0]{\spacefactor3000\relax}%
\providecommand \BibitemShut  [1]{\csname bibitem#1\endcsname}%
\let\auto@bib@innerbib\@empty
\bibitem [{\citenamefont {Evans}\ \emph {et~al.}(1993)\citenamefont {Evans},
  \citenamefont {Cohen},\ and\ \citenamefont {Morriss}}]{EvanCoheMorr93}%
  \BibitemOpen
  \bibfield  {author} {\bibinfo {author} {\bibfnamefont {D.~J.}\ \bibnamefont
  {Evans}}, \bibinfo {author} {\bibfnamefont {E.~G.~D.}\ \bibnamefont {Cohen}},
  \ and\ \bibinfo {author} {\bibfnamefont {G.~P.}\ \bibnamefont {Morriss}},\
  }\href@noop {} {\bibfield  {journal} {\bibinfo  {journal} {Phys. Rev. Lett.}\
  }\textbf {\bibinfo {volume} {71}},\ \bibinfo {pages} {2401} (\bibinfo {year}
  {1993})}\BibitemShut {NoStop}%
\bibitem [{\citenamefont {Evans}\ and\ \citenamefont
  {Searles}(1994)}]{EvanSear94}%
  \BibitemOpen
  \bibfield  {author} {\bibinfo {author} {\bibfnamefont {D.~J.}\ \bibnamefont
  {Evans}}\ and\ \bibinfo {author} {\bibfnamefont {D.~J.}\ \bibnamefont
  {Searles}},\ }\href@noop {} {\bibfield  {journal} {\bibinfo  {journal} {Phys.
  Rev. E}\ }\textbf {\bibinfo {volume} {50}},\ \bibinfo {pages} {1645}
  (\bibinfo {year} {1994})}\BibitemShut {NoStop}%
\bibitem [{\citenamefont {Gallavotti}\ and\ \citenamefont
  {Cohen}(1995)}]{GallCohe95a}%
  \BibitemOpen
  \bibfield  {author} {\bibinfo {author} {\bibfnamefont {G.}~\bibnamefont
  {Gallavotti}}\ and\ \bibinfo {author} {\bibfnamefont {E.~G.~D.}\ \bibnamefont
  {Cohen}},\ }\href@noop {} {\bibfield  {journal} {\bibinfo  {journal} {Phys.
  Rev. Lett.}\ }\textbf {\bibinfo {volume} {74}},\ \bibinfo {pages} {2694}
  (\bibinfo {year} {1995})}\BibitemShut {NoStop}%
\bibitem [{\citenamefont {Jarzynski}(1997{\natexlab{a}})}]{Jarz97PRL}%
  \BibitemOpen
  \bibfield  {author} {\bibinfo {author} {\bibfnamefont {C.}~\bibnamefont
  {Jarzynski}},\ }\href@noop {} {\bibfield  {journal} {\bibinfo  {journal}
  {Phys. Rev. Lett.}\ }\textbf {\bibinfo {volume} {78}},\ \bibinfo {pages}
  {2690} (\bibinfo {year} {1997}{\natexlab{a}})}\BibitemShut {NoStop}%
\bibitem [{\citenamefont {Jarzynski}(1997{\natexlab{b}})}]{Jarz97PRE}%
  \BibitemOpen
  \bibfield  {author} {\bibinfo {author} {\bibfnamefont {C.}~\bibnamefont
  {Jarzynski}},\ }\href@noop {} {\bibfield  {journal} {\bibinfo  {journal}
  {Phys. Rev. E}\ }\textbf {\bibinfo {volume} {56}},\ \bibinfo {pages} {5018}
  (\bibinfo {year} {1997}{\natexlab{b}})}\BibitemShut {NoStop}%
\bibitem [{\citenamefont {Crooks}(1998)}]{Crooks98}%
  \BibitemOpen
  \bibfield  {author} {\bibinfo {author} {\bibfnamefont {G.~E.}\ \bibnamefont
  {Crooks}},\ }\href@noop {} {\bibfield  {journal} {\bibinfo  {journal} {J.
  Stat. Phys.}\ }\textbf {\bibinfo {volume} {90}},\ \bibinfo {pages} {1481}
  (\bibinfo {year} {1998})}\BibitemShut {NoStop}%
\bibitem [{\citenamefont {Crooks}(1999)}]{Crooks99}%
  \BibitemOpen
  \bibfield  {author} {\bibinfo {author} {\bibfnamefont {G.~E.}\ \bibnamefont
  {Crooks}},\ }\href@noop {} {\bibfield  {journal} {\bibinfo  {journal} {Phys.
  Rev. E}\ }\textbf {\bibinfo {volume} {60}},\ \bibinfo {pages} {2721}
  (\bibinfo {year} {1999})}\BibitemShut {NoStop}%
\bibitem [{\citenamefont {Crooks}(2000)}]{Crooks00}%
  \BibitemOpen
  \bibfield  {author} {\bibinfo {author} {\bibfnamefont {G.~E.}\ \bibnamefont
  {Crooks}},\ }\href@noop {} {\bibfield  {journal} {\bibinfo  {journal} {Phys.
  Rev. E}\ }\textbf {\bibinfo {volume} {61}},\ \bibinfo {pages} {2361}
  (\bibinfo {year} {2000})}\BibitemShut {NoStop}%
\bibitem [{\citenamefont {Chetrite}\ \emph {et~al.}(2008)\citenamefont
  {Chetrite}, \citenamefont {Falkovich},\ and\ \citenamefont
  {Gawedzki}}]{ChetFalkGawe08}%
  \BibitemOpen
  \bibfield  {author} {\bibinfo {author} {\bibfnamefont {R.}~\bibnamefont
  {Chetrite}}, \bibinfo {author} {\bibfnamefont {G.}~\bibnamefont {Falkovich}},
  \ and\ \bibinfo {author} {\bibfnamefont {K.}~\bibnamefont {Gawedzki}},\
  }\href@noop {} {\bibfield  {journal} {\bibinfo  {journal} {J. Stat. Mech.
  Theor. Exp.}\ }\textbf {\bibinfo {volume} {2008}},\ \bibinfo {pages} {P08005}
  (\bibinfo {year} {2008})}\BibitemShut {NoStop}%
\bibitem [{\citenamefont {Garc\'ia-Garc\'ia}\ \emph {et~al.}(2012)\citenamefont
  {Garc\'ia-Garc\'ia}, \citenamefont {Lecomte}, \citenamefont {Kolton},\ and\
  \citenamefont {Dom\'inguez}}]{GarcLecoKoltDomi12}%
  \BibitemOpen
  \bibfield  {author} {\bibinfo {author} {\bibfnamefont {R.}~\bibnamefont
  {Garc\'ia-Garc\'ia}}, \bibinfo {author} {\bibfnamefont {V.}~\bibnamefont
  {Lecomte}}, \bibinfo {author} {\bibfnamefont {A.~B.}\ \bibnamefont {Kolton}},
  \ and\ \bibinfo {author} {\bibfnamefont {D.}~\bibnamefont {Dom\'inguez}},\
  }\href@noop {} {\bibfield  {journal} {\bibinfo  {journal} {J. Stat. Mech.
  Theor. Exp.}\ }\textbf {\bibinfo {volume} {2012}},\ \bibinfo {pages} {P02009}
  (\bibinfo {year} {2012})}\BibitemShut {NoStop}%
\bibitem [{\citenamefont {Wang}\ \emph {et~al.}(2002)\citenamefont {Wang},
  \citenamefont {Sevick}, \citenamefont {Mittag}, \citenamefont {Searles},\
  and\ \citenamefont {Evans}}]{WangSeviMittSear02}%
  \BibitemOpen
  \bibfield  {author} {\bibinfo {author} {\bibfnamefont {G.~M.}\ \bibnamefont
  {Wang}}, \bibinfo {author} {\bibfnamefont {E.~M.}\ \bibnamefont {Sevick}},
  \bibinfo {author} {\bibfnamefont {E.}~\bibnamefont {Mittag}}, \bibinfo
  {author} {\bibfnamefont {D.~J.}\ \bibnamefont {Searles}}, \ and\ \bibinfo
  {author} {\bibfnamefont {D.~J.}\ \bibnamefont {Evans}},\ }\href@noop {}
  {\bibfield  {journal} {\bibinfo  {journal} {Phys. Rev. Lett.}\ }\textbf
  {\bibinfo {volume} {89}},\ \bibinfo {pages} {050601} (\bibinfo {year}
  {2002})}\BibitemShut {NoStop}%
\bibitem [{\citenamefont {Carberry}\ \emph {et~al.}(2004)\citenamefont
  {Carberry}, \citenamefont {Williams}, \citenamefont {Wang}, \citenamefont
  {Sevick},\ and\ \citenamefont {Evans}}]{CarbWillWangSevi04}%
  \BibitemOpen
  \bibfield  {author} {\bibinfo {author} {\bibfnamefont {D.~M.}\ \bibnamefont
  {Carberry}}, \bibinfo {author} {\bibfnamefont {S.~R.}\ \bibnamefont
  {Williams}}, \bibinfo {author} {\bibfnamefont {G.~M.}\ \bibnamefont {Wang}},
  \bibinfo {author} {\bibfnamefont {E.~M.}\ \bibnamefont {Sevick}}, \ and\
  \bibinfo {author} {\bibfnamefont {D.~J.}\ \bibnamefont {Evans}},\ }\href@noop
  {} {\bibfield  {journal} {\bibinfo  {journal} {Jour. Chem. Phys.}\ }\textbf
  {\bibinfo {volume} {121}},\ \bibinfo {pages} {8179\textendash82} (\bibinfo
  {year} {2004})}\BibitemShut {NoStop}%
\bibitem [{\citenamefont {Garnier}\ and\ \citenamefont
  {Ciliberto}(2005)}]{GarnCili05}%
  \BibitemOpen
  \bibfield  {author} {\bibinfo {author} {\bibfnamefont {N.}~\bibnamefont
  {Garnier}}\ and\ \bibinfo {author} {\bibfnamefont {S.}~\bibnamefont
  {Ciliberto}},\ }\href@noop {} {\bibfield  {journal} {\bibinfo  {journal}
  {Phys. Rev. E}\ }\textbf {\bibinfo {volume} {71}},\ \bibinfo {pages}
  {060101(R)} (\bibinfo {year} {2005})}\BibitemShut {NoStop}%
\bibitem [{\citenamefont {Collin}\ \emph {et~al.}(2005)\citenamefont {Collin},
  \citenamefont {Ritort}, \citenamefont {Jarzynski}, \citenamefont {Smith},
  \citenamefont {Jr.},\ and\ \citenamefont {Bustamante}}]{CollRitoJarzSmit05}%
  \BibitemOpen
  \bibfield  {author} {\bibinfo {author} {\bibfnamefont {D.}~\bibnamefont
  {Collin}}, \bibinfo {author} {\bibfnamefont {F.}~\bibnamefont {Ritort}},
  \bibinfo {author} {\bibfnamefont {C.}~\bibnamefont {Jarzynski}}, \bibinfo
  {author} {\bibfnamefont {S.~B.}\ \bibnamefont {Smith}}, \bibinfo {author}
  {\bibfnamefont {I.}\ \bibnamefont {Tinoco,~Jr.}}, \ and\ \bibinfo {author}
  {\bibfnamefont {C.}~\bibnamefont {Bustamante}},\ }\href@noop {} {\bibfield
  {journal} {\bibinfo  {journal} {Nature}\ }\textbf {\bibinfo {volume} {437}},\
  \bibinfo {pages} {231\textendash4} (\bibinfo {year} {2005})}\BibitemShut
  {NoStop}%
\bibitem [{\citenamefont {Klenke}(2013)}]{Klenke13}%
  \BibitemOpen
  \bibfield  {author} {\bibinfo {author} {\bibfnamefont {A.}~\bibnamefont
  {Klenke}},\ }\href@noop {} {\emph {\bibinfo {title} {Probability Theory: A
  Comprehensive Course}}},\ \bibinfo {edition} {2nd}\ ed.,\ Universitext\
  (\bibinfo  {publisher} {Springer-Verlag},\ \bibinfo {address} {London, UK},\
  \bibinfo {year} {2013})\BibitemShut {NoStop}%
\bibitem [{\citenamefont {Maes}(1999)}]{Maes99}%
  \BibitemOpen
  \bibfield  {author} {\bibinfo {author} {\bibfnamefont {C.}~\bibnamefont
  {Maes}},\ }\href@noop {} {\bibfield  {journal} {\bibinfo  {journal} {J. Stat.
  Phys.}\ }\textbf {\bibinfo {volume} {95}},\ \bibinfo {pages} {367} (\bibinfo
  {year} {1999})}\BibitemShut {NoStop}%
\bibitem [{\citenamefont {Chetrite}\ and\ \citenamefont
  {Gupta}(2011)}]{ChetGupt11}%
  \BibitemOpen
  \bibfield  {author} {\bibinfo {author} {\bibfnamefont {R.}~\bibnamefont
  {Chetrite}}\ and\ \bibinfo {author} {\bibfnamefont {S.}~\bibnamefont
  {Gupta}},\ }\href@noop {} {\bibfield  {journal} {\bibinfo  {journal} {J.
  Stat. Phys.}\ }\textbf {\bibinfo {volume} {143}},\ \bibinfo {pages} {543}
  (\bibinfo {year} {2011})}\BibitemShut {NoStop}%
\bibitem [{\citenamefont {Seifert}(2005)}]{Seif05}%
  \BibitemOpen
  \bibfield  {author} {\bibinfo {author} {\bibfnamefont {U.}~\bibnamefont
  {Seifert}},\ }\href@noop {} {\bibfield  {journal} {\bibinfo  {journal} {Phys.
  Rev. Lett.}\ }\textbf {\bibinfo {volume} {95}},\ \bibinfo {pages} {040602}
  (\bibinfo {year} {2005})}\BibitemShut {NoStop}%
\bibitem [{\citenamefont {Esposito}\ and\ \citenamefont {{Van den
  Broeck}}(2010)}]{EspoBroe10}%
  \BibitemOpen
  \bibfield  {author} {\bibinfo {author} {\bibfnamefont {M.}~\bibnamefont
  {Esposito}}\ and\ \bibinfo {author} {\bibfnamefont {C.}~\bibnamefont {{Van
  den Broeck}}},\ }\href@noop {} {\bibfield  {journal} {\bibinfo  {journal}
  {Phys. Rev. Lett.}\ }\textbf {\bibinfo {volume} {104}},\ \bibinfo {pages}
  {090601} (\bibinfo {year} {2010})}\BibitemShut {NoStop}%
\bibitem [{\citenamefont {Seifert}(2012)}]{Seif12}%
  \BibitemOpen
  \bibfield  {author} {\bibinfo {author} {\bibfnamefont {U.}~\bibnamefont
  {Seifert}},\ }\href@noop {} {\bibfield  {journal} {\bibinfo  {journal} {Rep.
  Prog. Phys.}\ }\textbf {\bibinfo {volume} {75}},\ \bibinfo {pages} {126001}
  (\bibinfo {year} {2012})}\BibitemShut {NoStop}%
\bibitem [{\citenamefont {Aliprantis}\ and\ \citenamefont
  {Border}(2005)}]{AlipBord05}%
  \BibitemOpen
  \bibfield  {author} {\bibinfo {author} {\bibfnamefont {C.~D.}\ \bibnamefont
  {Aliprantis}}\ and\ \bibinfo {author} {\bibfnamefont {K.~C.}\ \bibnamefont
  {Border}},\ }\href@noop {} {\emph {\bibinfo {title} {Infinite Dimensional
  Analysis: A Hitchhiker's Guide}}},\ \bibinfo {edition} {3rd}\ ed.\ (\bibinfo
  {publisher} {Springer},\ \bibinfo {address} {Berlin},\ \bibinfo {year}
  {2005})\BibitemShut {NoStop}%
\bibitem [{\citenamefont {Ju}\ \emph {et~al.}(2005)\citenamefont {Ju},
  \citenamefont {Schaefer}, \citenamefont {Warren},\ and\ \citenamefont
  {Desbrun}}]{JuSchaWarrDesb05}%
  \BibitemOpen
  \bibfield  {author} {\bibinfo {author} {\bibfnamefont {T.}~\bibnamefont
  {Ju}}, \bibinfo {author} {\bibfnamefont {S.}~\bibnamefont {Schaefer}},
  \bibinfo {author} {\bibfnamefont {J.}~\bibnamefont {Warren}}, \ and\ \bibinfo
  {author} {\bibfnamefont {M.}~\bibnamefont {Desbrun}},\ }in\ \href@noop {}
  {\emph {\bibinfo {booktitle} {Eurographics Symposium on Geometry
  Processing}}},\ \bibinfo {editor} {edited by\ \bibinfo {editor}
  {\bibfnamefont {M.}~\bibnamefont {Desbrun}}\ and\ \bibinfo {editor}
  {\bibfnamefont {H.}~\bibnamefont {Pottmann}}}\ (\bibinfo  {publisher} {The
  Eurographics Association and Blackwell Publishing Ltd.},\ \bibinfo {address}
  {Oxford, UK},\ \bibinfo {year} {2005})\ pp.\ \bibinfo {pages}
  {181--6}\BibitemShut {NoStop}%
\bibitem [{\citenamefont {H\"ormander}(1990)}]{Horm90}%
  \BibitemOpen
  \bibfield  {author} {\bibinfo {author} {\bibfnamefont {L.}~\bibnamefont
  {H\"ormander}},\ }\href@noop {} {\emph {\bibinfo {title} {The Analysis of
  Linear Partial Differential Operators I: Distribution Theory and Fourier
  Analysis}}},\ \bibinfo {edition} {2nd}\ ed.\ (\bibinfo  {publisher}
  {Springer-Verlag},\ \bibinfo {address} {Berlin},\ \bibinfo {year}
  {1990})\BibitemShut {NoStop}%
\bibitem [{\citenamefont {Neri}\ \emph {et~al.}(2017)\citenamefont {Neri},
  \citenamefont {Rold\'an},\ and\ \citenamefont {J\"ulicher}}]{NeriRoldJuli17}%
  \BibitemOpen
  \bibfield  {author} {\bibinfo {author} {\bibfnamefont {I.}~\bibnamefont
  {Neri}}, \bibinfo {author} {\bibfnamefont {{\'E}.}~\bibnamefont {Rold\'an}},
  \ and\ \bibinfo {author} {\bibfnamefont {F.}~\bibnamefont {J\"ulicher}},\
  }\href@noop {} {\bibfield  {journal} {\bibinfo  {journal} {Phys. Rev. X}\
  }\textbf {\bibinfo {volume} {7}},\ \bibinfo {pages} {011019} (\bibinfo {year}
  {2017})}\BibitemShut {NoStop}%
\bibitem [{\citenamefont {Stratonovich}(1969)}]{Stra69}%
  \BibitemOpen
  \bibfield  {author} {\bibinfo {author} {\bibfnamefont {R.~L.}\ \bibnamefont
  {Stratonovich}},\ }\href@noop {} {\emph {\bibinfo {title} {Theory of Random
  Noise}}}\ (\bibinfo  {publisher} {Gordon and Breach},\ \bibinfo {address}
  {London},\ \bibinfo {year} {1969})\BibitemShut {NoStop}%
\bibitem [{\citenamefont {Gomez-Martin}\ and\ \citenamefont
  {Pagonabarraga}(2006)}]{MariPago06}%
  \BibitemOpen
  \bibfield  {author} {\bibinfo {author} {\bibfnamefont {A.}~\bibnamefont
  {Gomez-Martin}}\ and\ \bibinfo {author} {\bibfnamefont {I.}~\bibnamefont
  {Pagonabarraga}},\ }\href@noop {} {\bibfield  {journal} {\bibinfo  {journal}
  {Phys. Rev. E}\ }\textbf {\bibinfo {volume} {74}},\ \bibinfo {pages} {061113}
  (\bibinfo {year} {2006})}\BibitemShut {NoStop}%
\bibitem [{\citenamefont {Pietzonka}\ \emph {et~al.}(2016)\citenamefont
  {Pietzonka}, \citenamefont {Barato},\ and\ \citenamefont
  {Seifert}}]{PietBaraSeif16}%
  \BibitemOpen
  \bibfield  {author} {\bibinfo {author} {\bibfnamefont {P.}~\bibnamefont
  {Pietzonka}}, \bibinfo {author} {\bibfnamefont {A.~C.}\ \bibnamefont
  {Barato}}, \ and\ \bibinfo {author} {\bibfnamefont {U.}~\bibnamefont
  {Seifert}},\ }\href@noop {} {\bibfield  {journal} {\bibinfo  {journal} {Phys.
  Rev. E}\ }\textbf {\bibinfo {volume} {93}},\ \bibinfo {pages} {052145}
  (\bibinfo {year} {2016})}\BibitemShut {NoStop}%
\bibitem [{\citenamefont {Ginrich}\ \emph {et~al.}(2016)\citenamefont
  {Ginrich}, \citenamefont {Horowitz}, \citenamefont {Perunov},\ and\
  \citenamefont {England}}]{GingHoroPeruEngl16}%
  \BibitemOpen
  \bibfield  {author} {\bibinfo {author} {\bibfnamefont {T.~R.}\ \bibnamefont
  {Ginrich}}, \bibinfo {author} {\bibfnamefont {J.~M.}\ \bibnamefont
  {Horowitz}}, \bibinfo {author} {\bibfnamefont {N.}~\bibnamefont {Perunov}}, \
  and\ \bibinfo {author} {\bibfnamefont {J.}~\bibnamefont {England}},\
  }\href@noop {} {\bibfield  {journal} {\bibinfo  {journal} {Phys. Rev. Lett.}\
  }\textbf {\bibinfo {volume} {116}},\ \bibinfo {pages} {120601} (\bibinfo
  {year} {2016})}\BibitemShut {NoStop}%
\bibitem [{\citenamefont {Polettini}\ \emph {et~al.}(2016)\citenamefont
  {Polettini}, \citenamefont {Lazarescu},\ and\ \citenamefont
  {Esposito}}]{PoleLazaEspo16}%
  \BibitemOpen
  \bibfield  {author} {\bibinfo {author} {\bibfnamefont {M.}~\bibnamefont
  {Polettini}}, \bibinfo {author} {\bibfnamefont {A.}~\bibnamefont
  {Lazarescu}}, \ and\ \bibinfo {author} {\bibfnamefont {M.}~\bibnamefont
  {Esposito}},\ }\href@noop {} {\bibfield  {journal} {\bibinfo  {journal}
  {Phys. Rev. E}\ }\textbf {\bibinfo {volume} {94}},\ \bibinfo {pages} {052104}
  (\bibinfo {year} {2016})}\BibitemShut {NoStop}%
\bibitem [{\citenamefont {Lebowitz}\ and\ \citenamefont
  {Spohn}(1999)}]{LeboSpoh98}%
  \BibitemOpen
  \bibfield  {author} {\bibinfo {author} {\bibfnamefont {J.~L.}\ \bibnamefont
  {Lebowitz}}\ and\ \bibinfo {author} {\bibfnamefont {H.}~\bibnamefont
  {Spohn}},\ }\href@noop {} {\bibfield  {journal} {\bibinfo  {journal} {J.
  Stat. Phys.}\ }\textbf {\bibinfo {volume} {95}},\ \bibinfo {pages} {333}
  (\bibinfo {year} {1999})}\BibitemShut {NoStop}%
\end{thebibliography}
%

\end{document}